\documentclass{article}
\usepackage{algorithm}
\usepackage{algpseudocode}

\usepackage{amsmath}
\usepackage{amssymb}

\usepackage{enumitem}
\usepackage[english]{babel}
\usepackage[utf8]{inputenc}
\usepackage{amsmath,amssymb}
\usepackage{graphicx,caption,subcaption,multirow}
\usepackage[colorinlistoftodos]{todonotes}

\usepackage{multirow}
\usepackage[T1]{fontenc}
\usepackage[utf8]{inputenc}
\usepackage{authblk}
\usepackage{xr}

\newcommand{\bm}{{\bf m}}

\newcommand{\bG}{{\mathbb{G}}}
\newcommand{\bEg}{{\mathbb{E}}}

\newcommand{\cC}{\mathcal{C}}

\newcommand{\cS}{\mathcal{S}}
\newcommand{\cN}{\mathcal{N}}

\newcommand{\bP}[1]{{{\bf P}}\left[{#1}\right]}

\newcommand{\bE}[1]{{\mathbb{E}}\left[{#1}\right]}
\newcommand{\1}[1]{{\bf 1}\left[#1\right]}

\newcommand{\eat}[1]{}

\newtheorem{theorem}{Theorem}[section]

\newtheorem{lemma}[theorem]{Lemma}

\newcommand{\fsquare}{\vrule height6pt width7pt depth1pt}   
\newcommand{\myproof}{{\hfill \\ \bf Proof. \ }}           
\newcommand{\myendpf}{\hfill\fsquare \\[0.1in]}

\begin{document}
%
\title{An efficient alternative to Ollivier-Ricci curvature based on the Jaccard metric
\thanks{Research was sponsored by the Army Research Laboratory and was accomplished under Cooperative Agreement Number W911NF-09-2-0053 (the ARL Network Science CTA). The views and conclusions contained in this document are those of the authors and should not be interpreted as representing the official policies, either expressed or implied, of the Army Research Laboratory or the U.S. Government. The U.S. Government is authorized to reproduce and distribute reprints for Government purposes notwithstanding any copyright notation here on. This document does not contain technology or technical data controlled under either the U.S.
International Traffic in Arms Regulations or the U.S. Export
Administration Regulations. }}

\author[1]{Siddharth Pal\thanks{siddharth.pal@raytheon.com}}
\author[2]{Feng Yu\thanks{fyu@gc.cuny.edu}}
\author[3]{Terrence J. Moore\thanks{terrence.j.moore.civ@mail.mil}}
\author[4]{Ram Ramanathan\thanks{Ram@gotenna.com}
\thanks{Work done while the author was with Raytheon BBN Technologies}}
\author[2]{Amotz Bar-Noy\thanks{amotz@sci.brooklyn.cuny.edu}}
\author[3]{Ananthram Swami\thanks{ananthram.swami.civ@mail.mil}}
\affil[1]{Raytheon BBN Technologies, Cambridge, MA 02138, USA.}
\affil[2]{Graduate Center of the City University of New York, New York, NY 10016, USA.}
\affil[3]{U.S. Army Research Lab, Adelphi, MD 20783, USA.}
\affil[4]{GoTenna Inc, Brooklyn, NY 11201, USA.}
\renewcommand\Authands{ and }





%


\maketitle



\newcommand{\SC}{\ensuremath{\Delta}}
\newcommand{\NB}{\ensuremath{\mathcal{N}}}

\begin{abstract}
We study Ollivier-Ricci curvature, a discrete version of Ricci curvature, which has gained popularity over the past several years and has found applications in diverse fields. However, the Ollivier-Ricci curvature requires an optimal mass transport problem to be solved, which can be computationally expensive for large networks. In view of this, we propose two alternative measures of curvature to Ollivier-Ricci  which are motivated by the Jaccard coefficient and are demonstrably less computationally intensive, a cheaper Jaccard (JC) and a more expensive generalized Jaccard (gJC) curvature metric. We show theoretically that the gJC closely matches the Ollivier-Ricci curvature for Erd{\"o}s-R{\'e}nyi graphs in the asymptotic regime of large networks. Furthermore, we study the goodness of approximation between the proposed curvature metrics and Ollivier-Ricci curvature for several network models and real networks. Our results suggest that in comparison to an alternative curvature metric for graphs, the Forman-Ricci curvature, the gJC exhibits a reasonably good fit to the Ollivier-Ricci curvature for a wide range of networks, while the JC is shown to be a good proxy only for certain scenarios.

\end{abstract}

\section{Introduction} 

The various notions of curvature in differential geometry measure, in different ways, the curves or bends of tensors on the surface of a manifold~\cite{jost2008riemannian, gallot1987riemannian, goldberg1998curvature}. Several of these definitions of curvature have recently been interpreted on graphs and applied to networks. Some examples include Gaussian curvature~\cite{higuchi2001combinatorial}, Gromov curvature \cite{narayan2011large}, and Ricci curvature~\cite{forman2003bochner, ollivier2009ricci}. Of these, the Ollivier-Ricci curvature seems to be the most promising new metric for networks. It has been shown to be able to measure robustness in gene expression, reliably distinguishing between cancerous and non-cancerous cells~\cite{sandhu2015graph}. Ricci curvature has been shown to indicate fragility in stock markets~\cite{sandhu2016ricci}. Ollivier-Ricci curvature has also been applied to explaining congestion wireless network capacity~\cite{wang2014wireless}.

Ollivier-Ricci curvature is defined between a pair of vertices in a network based on the optimal mass transport, determined by the Wasserstein distance, between their associated mass distributions. When restricted to the transport between adjacent vertices, Ollivier-Ricci curvature can be viewed as a edge centrality metric, akin to betweenness or random-walk measures on edges. Positive curvature implies the neighbors of the two nodes are close (perhaps overlapping or shared). Zero (or near-zero) curvature implies the nodes are locally embeddable in a flat surface (as in a grid or regular lattice). Negative curvature implies that the neighbors of the two nodes are further apart. 

Unfortunately, Ollivier-Ricci curvature can have high-computational complexity in dense, high-degree networks as solving the Wasserstein distance can, in the worst case, scale with the quartic of the degree (see Sec. \ref{sec:ComputationalComplexity}) or, in practice, scale with the product of the two nodes' degrees~\cite{ni2015ricci}. This motivates the desire for a less computationally-intensive approximation. Jost and Liu~\cite{jost2014ollivier} demonstrated the significance of overlapping neighborhoods in the Ollivier-Ricci curvature of edges in the formulation of a bound involving the clustering coefficient~\cite{watts1998collective}. Hence, it seems reasonable to build a metric approximating Ollivier-Ricci from the sets of common and separate neighbors of the nodes in an edge.

We derive a new curvature metric approximating the Ollivier-Ricci graph curvature metric using the Jaccard index, which has previously found utility in networks, e.g., as a measure of similarity between nodes~\cite{liben2007link}. The Jaccard index naturally captures the overlapping neighborhood feature found in positively curved edges in a simplistic manner. The notion of set-comparison as a curvature metric leads to a more general linear approximation function of Ollivier-Ricci formulated from classes of sets of each node's neighbors that effectively solves a mass exchange problem. The complexity of this new metric is significantly less than that for Ollivier-Ricci. For random graphs, we find that our new metric shares many asymptotic properties of the Ollivier-Ricci curvature~\cite{bhattacharya2015exact}. Moreover, comparisons of the Jaccard-inspired curvature with Ollivier-Ricci seem more favorable than the alternative Forman-Ricci curvature metric~\cite{forman2003bochner, sreejith2016forman, sreejith2017systematic} that is extremely computationally efficient, on network models and real networks.

\externaldocument{ms.tex}
\externaldocument{experimental_results.tex}
\externaldocument{theoretical_results.tex}

\section{Curvature Metrics}
\label{sec:curvature_defns}

We first introduce two discretized version of the Ricci curvature --  the Ollivier-Ricci curvature as discussed in \cite{ollivier2009ricci,jost2014ollivier,lin2011ricci}, and the Forman-Ricci curvature as discussed in \cite{forman2003bochner,sreejith2016forman}. Then, we introduce a new graph curvature metric which is intuitively similar to Ollivier-Ricci curvature, without requiring as much computational complexity. 
 
\subsection{Ollivier-Ricci curvature}
\label{subsec:ORcurvature}
Consider an undirected graph $G=(V,E)$ on $n$ nodes, i.e., $|V|=n$, with no self loops. We define the metric $d$ such that for distinct vertices $i$,$j$, $d(i,j)$ is the length of the shortest path connecting $i$ and $j$.  Ollivier-Ricci curvature can be defined on the graph $G$, with a probability measure $\bm _i$ attached to each vertex $i \in V$. For two nodes $i$ and $j$, we define a mass transport plan $\nu_{i,j}: V \times V \to [0,1]$
such that for every $x$, $y$ in $V$
\begin{equation}
\sum _{k \in V} \nu_{i,j}(x,k) = m _i(x) \ \mbox{ and } 
\sum _{\ell \in V} \nu_{i,j}(\ell,y) = m_j(y).
\end{equation}  
The above condition enforces that mass attributed to any neighbor of $i$, $m_i(\cdot)$, is completely transferred to neighbors of $j$ in such a manner that all the neighbors of $j$ get exactly
the required mass $m_j(\cdot)$.
 Let the space of all valid mass transport plans between nodes $i$ and $j$ be denoted by $\Pi(i,j)$.

For an edge $(i,j) \in E$, the Ollivier-Ricci (OR) curvature metric is defined as follows
\begin{equation}
\kappa (i,j) = 1 - W(\bm _i, \bm _j),
\label{eq:OR-Definition}
\end{equation}
where $W(\bm _i, \bm _j)$ is the Wasserstein distance or optimal mass transport cost between the two probability measures $\bm_i$ and $\bm_j$, expressed as follows
\begin{equation}
W(\bm_i,\bm_j) = \inf _{\nu_{i,j} \in \Pi(i,j)} \sum _{x \in V} \sum _{y \in V} 
\nu_{i,j}(x,y) d(x,y).
\end{equation}
For each vertex $i$, the probability measure $\bm _i$ is set as
\begin{align}
m_i(j) &= \frac{1}{d_i}, \ \mbox{if } i \sim j
\nonumber \\
&= 0 \ \mbox{ otherwise,}
\end{align}
where $i \sim j$ implies an edge between $i$ and $j$. 
The probability measure $\bm_i$ shown above assigns weight to all neighbors of $i$
uniformly as in \cite{sandhu2015graph,jost2014ollivier}. A more generic setting is where a mass $0 \leq \alpha \leq 1$ is assigned to node $i$, and the rest of the mass $1-\alpha$ is distributed uniformly among the neighbors of $i$~\cite{ni2015ricci,lin2011ricci}. 

We can bound the Ollivier-Ricci curvatures defined in \eqref{eq:OR-Definition}. First, note that for an edge $(i,j)$, the minimum distance between a neighbor of $i$ and $j$ is $0$ when they are common, and the maximum distance is $3$ hops. This implies the following bound on the Wasserstein distance $0< W(\bm_i,\bm_j) <3$, which in turn implies 
$-2 < \kappa < 1$. 

\subsection{Forman-Ricci curvature}

Forman discretized the classical Ricci curvature for a broad class of geometric objects, the CW complexes~\cite{forman2003bochner}, which is called the Forman-Ricci or simply Forman curvature. While the original definition of Forman curvature for CW complexes is not relevant to this paper, we present Forman curvature for undirected networks as introduced in~\cite{sreejith2016forman}. This was proposed as a candidate for a discrete Ricci curvature to gain new insights on the organization of complex networks. 

The Forman curvature for an edge $e=(i,j)$ is given as follows
\begin{equation}
F(e) = w(e) \left( \frac{w_i}{w_e} + \frac{w_j}{w_e} - 
\sum _{e_\ell \in e_{i} \setminus e} \frac{w_i}{\sqrt{w_e w_{e_\ell}}}
- \sum _{e_\ell \in e_{j} \setminus e}  \frac{w_j}{\sqrt{w_e w_{e_\ell}}}
\right)
\label{eq:Forman_complete_defn}
\end{equation}
where $w_e$ is a weight associated with the edge $e$, $w_i$ and $w_j$ are weights associated with vertices $i$ and $j$, and $e_i \setminus e$ and $e_j \setminus e$ denote the set of edges incident on vertices $i$ and $j$ excluding the edge $e$. 

For an unweighted graph, two weighting schemes were proposed~\cite{sreejith2016forman,sreejith2017systematic,saucan2017discrete}. One was to set all the node and edge weights as $1$, and the other was to weight the edges by $1$, and the nodes by their degree. We implemented both the weighting schemes, and did not find a significant difference in terms of their correlations with the Ollivier-Ricci curvature. The results shown in Section \ref{sec:experimental_results} follow the former weighting scheme, where the original expression of Forman curvature \eqref{eq:Forman_complete_defn} reduces to
\begin{equation}
F(e) = 4 - d_i - d_j.
\label{eq:FormanCurvature_weighted_expression}
\end{equation} 
Other, more involved, weighting schemes have also been 
proposed~\cite{weber2017characterizing,weber2016forman}, but are not considered in this work.

\subsection{Jaccard Curvatures}
\label{subsec:JCcurvature}

Calculating Ollivier-Ricci curvature can be costly because it involves solving an optimal mass transport problem, or equivalently a linear program~\cite{ni2015ricci}, for each edge. Especially for large graphs, with high values of maximum degree, calculating OR curvature for all the edges can be prohibitively costly (see Section \ref{sec:ComputationalComplexity}). To address this issue, we introduce an approximation to the OR curvature, which would not require solving the optimal mass transport problem. Towards this end, we revisit the intuition of OR curvature -- An edge has positive curvature if the neighborhoods of the two concerned nodes are closer to each other compared to the nodes themselves, zero curvature if the neighborhoods are at the same distance, and negative curvature if the neighborhoods are farther apart. A simple heuristic would be to measure the fraction of common nodes between the neighborhoods of the two concerned nodes. This is related to Jaccard's coefficient~\cite{liben2007link} which was introduced in network analysis as a similarity measure between nodes.
First, we define some notation:
For an edge $(i,j)$, the set of common neighbors of the nodes $i$ and $j$ is given by
\[
\cC(i,j) = \cN_i \cap \cN_j,
\]
where $\cN _k$ is the neighbor set of node $k$. We let $C(i,j) = | \cC(i,j) |$. We also define the set of separate neighbors between $i$ and $j$ as follows
\[
\cS(i,j) = \left( \cN_i \cup \cN_j \right) \setminus \cC(i,j),
\]
with $S(i,j) = | \cS(i,j) |$. We define the union of the neighbor sets of $i$ and $j$ as $\cN(i,j)$, i.e.,
\[
\cN(i,j) = \cC(i,j)  \cup \cS(i,j),
\]
with $N(i,j) = |\cN(i,j)|$. 
 
\begin{figure}[h]
 \centering
  \includegraphics[width=0.5\textwidth]{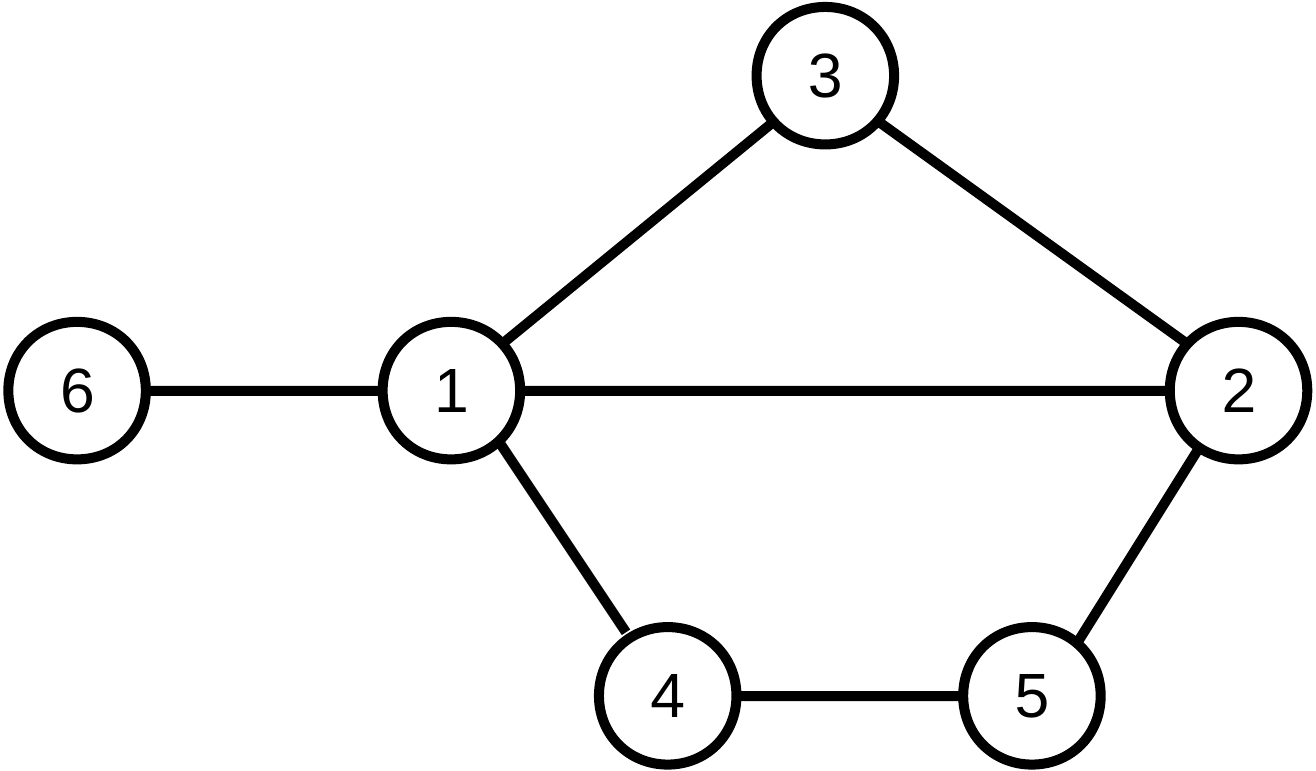}
 \caption{Jaccard curvatures shown for an illustrative example -- We consider the edge $(1,2)$ and calculate its Jaccard (JC), see \eqref{eq:JCA_another_expression}, and generalized Jaccard (gJC) curvatures, see \eqref{eq:mJCA_finalExpression} for details. We have one common neighbor, i.e., $C(1,2) = \{3\}$, and the set of separate numbers is $\{1,2,4,5,6\}$. Therefore, $C(1,2)=1$, $S(1,2) = 5$ and $N(1,2) = 6$, and the Jaccard curvature $JC(1,2) = 1- \frac{3\times 5}{6} =  -\frac{3}{2}$. Since, node $4$ is the only exclusive neighbor of node $1$ directly connected to an exclusive neighbor of node $2$, we have $\cS_1 ^{(1)} = \{4\}$. By the same argument, we have $\cS_2 ^{(1)} = \{5\}$. The remaining exclusive neighbor of node $1$, i.e., node $6$, is connected to node $3$ (an exclusive neighbor of node $2$) through a path of $2$ hops, and $\cS_1 ^{(2)} = \{6\}$. Putting all this together, using \eqref{eq:mJCA_finalExpression} we obtain, $gJC(1,2) = 1 - \frac{1+1+2}{6} - 2 \cdot \frac{1}{6} = 0$. In comparison, the Ollivier-Riccci curvature $OR(1,2) = \frac{1}{4}$. }
\label{fig:Jaccard_curvature_example}
 \end{figure}

Jaccard's coefficient is defined as the ratio between the intersection of neighborhoods of the two nodes to their union, i.e.,
\begin{equation}
J(i,j) = \frac{C(i,j)}{N(i,j)}.
\label{eq:Jaccard_metric}
\end{equation}
It is evident that the metric $J(i,j)$ will be closer to $1$ if there are more common nodes, and closer to $0$ otherwise. However, the range of this metric will be between 0 and 1, as opposed to OR curvature which takes the range $(-2,1)$. We want the Jaccard curvature metric for an edge $(i,j)$, $JC(i,j)$, to approach a value of 1 when the fraction of common nodes to total nodes is close to $1$, and $-2$ when that fraction is equal to $0$. In other words, we have the following requirements,
\begin{align}
\mbox{When } \frac{C(i,j)}{N(i,j)} \approx 1, \ \mbox{then } JC(i,j) \approx 1
\label{eq:Jaccard_curvature_eq1}
\end{align}
and
\begin{align}
\mbox{when } \frac{C(i,j)}{N(i,j)} =0, \ \mbox{then } JC(i,j) = -2.
\label{eq:Jaccard_curvature_eq2}
\end{align}
Equations~\eqref{eq:Jaccard_curvature_eq1} and \eqref{eq:Jaccard_curvature_eq2} lead to the following expression for the Jaccard curvature, 
\begin{align}
JC(i,j) = 1 - \frac{3 S(i,j)}{N(i,j)} = -2 + 3 J(i,j).
\label{eq:JCA_another_expression}
\end{align}
see Figure~\ref{fig:Jaccard_curvature_example} for an illustrative example.

\noindent
The above expression could be interpreted as subtracting the influence of separate neighbors, with 
$S(i,j)$ being the total number of separate neighbors and the denominator $N(i,j)$ being the cardinality of the union of the neighbor sets of $i$ and $j$. 

Computing Jaccard curvature is very cheap because it only requires the knowledge of the size of neighborhoods of the two relevant nodes and the common nodes in those neighborhood sets. From simulations and experiments reported in Section \ref{sec:experimental_results}, we have observed that the Jaccard curvature is a reasonably good approximation of the Ollivier-Ricci curvature for several instances of generated and real-world networks. However, since the Jaccard curvature partitions the set of neighbors into common and separate vertices, the granularities of OR curvature is lost to a great extent. This is best demonstrated by considering randomly chosen edges in canonical graphs.

In a complete graph the OR curvature of each edge will be close to 1, and the Jaccard curvature will be exactly 1. For an edge connecting high degree nodes in a tree, the OR curvature will be close to -2, while the Jaccard curvature will be exactly -2. However on a grid or a line, the OR curvature of the edges will be 0, while the Jaccard curvature will still be -2, because there are no common nodes. Clearly, the Jaccard curvature metric should have a more positive value in a grid compared to that on a tree, if we are to obtain a better approximation of the OR curvature. To address this issue, we now define a generalized version of the Jaccard curvature metric to take into account nodes that are not common, yet closer than 3 hops apart.  

We introduce some more notation:
Define $\cN_i(i,j)$ as the exclusive neighbors of $i$ with respect to the edge $(i,j)$, i.e.,
\[
\cN_i(i,j) = \{ k \in V \setminus \{j\} \ | \ (i,k) \in E \}.
\]
For any two nodes $u$ and $v$, recall that $d(u,v)$ denotes the shortest path length between the
two nodes. 
Let the set of separate nodes be partitioned into the following sets
\[
\cS^{(r)} _i = \{ k \in \cN _i(i,j) \ | \ \min _{\ell \in \cN _j(i,j) } d(k,\ell) = r \},
\]
with $S_i ^{(r)} = |\cS_i ^{(r)}|$,
for $r=1,2,3$.  
In other words, $\cS_i ^{(r)}$ is the set of neighbors of $i$ that are at a distance of $r$ hops from the closest exclusive neighbor of $j$. 
If $\cN_j(i,j) = \emptyset$, then we set $\cS_i ^{(1)} = \cN_i(i,j)$.
Observe that
\[
\cS(i,j) = 
\left( \cup _{r =1} ^3 \cS^{(r)}_i \right) \cup 
\left( \cup _{r =1} ^3 \cS^{(r)}_j \right)  \cup \{i \cup j\}.
\]

Since Ollivier-Ricci curvature includes the endpoints of the edge itself, we include $i$ and $j$ in the generalized Jaccard (gJC) metric. 

Therefore, for an edge $(i,j) \in E$, the generalized Jaccard metric is defined as follows
\begin{equation}
gJC (\alpha,\beta,\gamma,\delta,\zeta;i,j) = \alpha + \beta \frac{C(i,j)}{N(i,j)} + \gamma \frac{S^{(1)}_i + S^{(1)}_j + 2}{N(i,j)}
+ \delta \frac{S^{(2)}_i + S^{(2)}_j}{N(i,j)}
+ \zeta \frac{S^{(3)}_i + S^{(3)}_j}{N(i,j)}
\label{eq:gJC_expression}
\end{equation}
where the parameters $\alpha$,$\beta$, $\gamma$, $\delta$ and $\zeta$ need to be determined. 
Since, $i \notin \cS^{(1)}_j$ and $j \notin \cS^{(1)}_i$, we arrive at the gJC metric by including the two nodes separately in \eqref{eq:gJC_expression}. 
The parameters are determined by considering the cases of canonical graphs. 

\subsubsection{Considering canonical graphs}

In a $k$-complete graph, we would like the generalized Jaccard metric to have a maximum value close to 1, so as to approximate the OR curvature which approaches $1$ as $k$ gets large.
Therefore we require, as $k\to \infty$ and
\begin{equation}
\frac{C(i,j)}{N(i,j)} \to 1, \ \mbox{then} \ gJC(i,j) \to 1,
\nonumber
\end{equation}
which leads to
\begin{equation}
\alpha + \beta = 1.
\label{eq:MJCA_eq1}
\end{equation}
Similarly, for edges in a $d$-dimensional grid, we would like the generalized Jaccard metric to have a value close to 0 as $d$ gets large. This requires that as $d \to \infty$ and
\begin{equation}
\frac{S^{(1)}_i + S^{(1)}_j + 2}{N(i,j)} \to 1, \ \mbox{then} \ gJC(i,j) \to 0,
\nonumber
\end{equation}
which leads to
\begin{equation}
\alpha + \gamma = 0.
\label{eq:mJCA_eq2}
\end{equation}
This would best approximate the OR curvature which has a value of $0$ for $d$ -dimensional grids.

For edges in a tree connecting nodes with degree $d$, we would like the generalized Jaccard metric to have a value close to a minimum value of -2, approximating the OR curvature which itself approaches 
$-2$ as $d$ gets large. 
This requires that as $d \to \infty$ and
\begin{equation}
\frac{S^{(3)}_i + S^{(3)}_j}{N(i,j)} \to 1, \ \mbox{then} \ gJC(i,j) \to -2
\nonumber
\end{equation}
which leads to
\begin{equation}
\alpha + \zeta = -2.
\label{eq:mJCA_eq3}
\end{equation}
We set, $\alpha = 1$, $\beta = 0$, $\gamma = -1$, $\zeta = -3$, which satisfies \eqref{eq:MJCA_eq1}-\eqref{eq:mJCA_eq3}. 
We also enforce the following bound
\begin{equation}
\gamma > \delta > \zeta.
\end{equation}
so that the effect of $\cS^{(2)}$ on the edge curvature 
falls between that of $\cS^{(1)}$ and $\cS^{(3)}$. 
Setting $\delta = 2$ or $\frac{\gamma+\zeta}{2}$, we obtain
\begin{equation}
gJC(i,j) = 1 -  \frac{S^{(1)}_i + S^{(1)}_j+2}{N(i,j)}
- 2 \frac{S^{(2)}_i + S^{(2)}_j}{N (i,j)}
- 3 \frac{S^{(3)}_i + S^{(3)}_j}{N (i,j)}.
\label{eq:mJCA_finalExpression}
\end{equation}

\subsubsection{Connections to mass transport}

The Ollivier-Ricci curvature is related to the solution of an optimal mass transport problem as discussed in Section 
\ref{subsec:ORcurvature}. Here we study if the Jaccard curvature has a similar connection to mass transport. Observe that the generalized Jaccard expression derived in \eqref{eq:mJCA_finalExpression} subtracts the influence of a node in $\cS^{(1)}$ with a weight of $1$, influence of a node in $\cS^{(2)}$ with a weight of $2$, and similarly for a node in $\cS^{(3)}$ with a weight of $3$. These weights exactly match the transportation cost of moving mass from a source node to any destination node in the neighborhood of the other node. Next, we observe that the Jaccard curvature is related to the solution of this optimal mass exchange problem. 

We define a mass exchange problem with an initial mass distribution of $m_k = \frac{1}{N(i,j)}$ for every $k$ in $\cN(i,j)$. The mass exchange plan $\nu _{i,j} : V \times V \to [0,1]$ requires that for every $x$ in $\cN_i(i,j)$ and $y$ in $\cN_j(i,j)$
\begin{align}
 \sum _{\ell \in V} \nu _{i,j}(x, \ell) = \frac{1}{N(i,j)}
\mbox{ and } 
\sum _{k \in V}  \nu _{i,j}(k, y) = \frac{1}{N(i,j)}
\label{eq:MassExchange_Defn}
\end{align}
and $\nu_{i,j}(i,j)= \frac{1}{N(i,j)}$ and $\nu_{i,j}(j,i)= \frac{1}{N(i,j)}$. The mass exchange problem introduced here only requires that mass from a particular node $x$ in $\cN_i(i,j)$ be completely transported to $\cN_j(i,j)$ and vice-versa, along with the requirement that mass at node $i$ be transported to node $j$ and vice-versa.

It can be shown that the generalized Jaccard expression in \eqref{eq:mJCA_finalExpression} is related to the solution of the optimal mass exchange problem between neighborhoods of the two concerned nodes, where the mass distribution at the source is predetermined and fixed and the destination mass distribution is kept flexible, with the constraint that mass from a neighbor of one node needs to be transported to any neighbor of the other node and vice-versa.
\externaldocument{ms.tex}
\externaldocument{curvature_setting.tex}
\externaldocument{experimental_results.tex}

\section{Analytical results on random graphs}
\label{sec:theory}

We state the following result on the behavior of gJC curvature in ER graphs and compare with that of the OR curvature \cite{bhattacharya2015exact}. We present the results for a sequence of Erdos-Renyi graphs $\{ \bG_1, \bG_2, \ldots\}$, 
and let $JC_n(i,j)$ and $gJC_n(i,j)$ denote the Jaccard and generalized Jaccard curvature of edge $(i,j)$ in the graph $\bG_n$. 

\begin{theorem}
{
Let $\{\bG_1,\bG_2, \ldots\}$ be a sequence of Erdos-Renyi graphs. 
As $n\to \infty$ and for all $(i,j) \in \bEg$, we have the following results.
\begin{enumerate}[label=\alph*.]
\item
For $p_n \to p$
\begin{equation}
\bE{JC_n(i,j)} \to \frac{5p-4}{2-p}
\label{eq:JC_lemma_ER_p_fixed}.
\end{equation}
\item
For $p_n \to 0$, $\bE{JC_n(i,j)} \to -2$ .
\end{enumerate}
\label{thm:MainTheorem-JC}
}
\end{theorem}

\begin{theorem}
{
Let $\{\bG_1,\bG_2, \ldots\}$ be a sequence of Erdos-Renyi graphs. 
As $n\to \infty$ and for all $(i,j) \in \bEg$, we have the following results.
\begin{enumerate}[label=\alph*.]
\item
For $p_n \to p$
\begin{equation}
\bE{gJC_n(i,j)} \to \frac{p}{2-p} .
\label{eq:lemma_ER_p_fixed}
\end{equation}
Note that $\bE{gJC_n} >0$ for all $p>0$. As $p \to 1$, $\bE{gJC_n(i,j)} \to 1$ and as $p \to 0$, $\bE{gJC_n(i,j)} \to 0$.
\item
For $n p_n \to 0$ and $p_n \to 0$, $\bE{gJC_n(i,j)} \to 0$.
\item
For $n  p_n ^2 \to \infty$ and $p_n \to 0$, $\bE{gJC_n(i,j)} \to 0$.
\item
For $n ^2 p_n ^3 \to \infty$, $n p_n ^2 \to 0$ and $p_n \to 0$, $\bE{gJC_n(i,j)} \to -1$.
\item
For $n p_n \to \infty$,$n^2 p _n ^3 \to 0$ and $p_n \to 0$, $\bE{gJC_n(i,j)} \to -2$.
\end{enumerate}
\label{thm:MainTheorem}
}
\end{theorem}

\noindent
\textbf{Proof of Theorems \ref{thm:MainTheorem-JC} and \ref{thm:MainTheorem}.}

Without loss of generality, we set $i=1$ and $j=2$ in \eqref{eq:mJCA_finalExpression}. 
Note that for the generic edge $(1,2)$ in graph $\bG_n$,
\[
N _n(1,2) = \sum  _{k \in V_n} \left[  \1{ k \sim 1, k \sim 2 } + \1{ k \sim 1, k \nsim 2 }
+ \1{ k \nsim 1, k \sim 2 } \right]. 
\]
Therefore, 
\begin{equation}
\bE{N _n (1,2)} = (n-2) p_n ^2 + \left[ 2 (n-2) p_n (1 - p_n ) +2 \right]
\label{eq:N_n_behavior}
\end{equation}
where, 
\begin{equation}
\bE{C_n (1,2)} = (n-2) p _n ^2
\label{eq:C_n_behavior}
\end{equation}
and 
\begin{equation}
\bE{S _n(1,2)} = 2 (n-2) p_n (1 - p_n )+2.
\label{eq:SeparateVertices_asymp}
\end{equation}
Observe that the number of common and separate nodes $C_n(1,2)$,$S_n(1,2)$ have been indexed by $n$, to denote that they correspond to the graph $\bG_n$.
Note that as $n \to \infty$ and if $p _n \to 0$, $\bE{S_n(1,2)} \sim 2 n p_n$ and 
$\bE{C_n(1,2)} \sim o(n p_n)$. However, if $p_n \to p$, then $\bE{S_n(1,2)} \to 2n p (1-p)$
and $\bE{C_n(1,2)} \to n p^2$.

For $(i,j) \in E$, the Jaccard curvature for the regime $p_n \to p$ follows by simply using \eqref{eq:N_n_behavior}-\eqref{eq:SeparateVertices_asymp}. For the other regimes where $p_n \to 0$, the fraction of common nodes to the neighborhood size goes to $0$, implying that $JC(i,j) \to -2$. This proves Theorem~\ref{thm:MainTheorem-JC}.

The argument for the gJC curvature is a bit more involved. From \eqref{eq:mJCA_finalExpression}, we have 
\begin{equation}
\bE{gJC(i,j)} = 1  - \bE{ \frac{S^{(1)}_i + S^{(1)}_j+2}{N (i,j)} }
 - 2 \bE{ \frac{S^{(2)}_i + S^{(2)}_j}{N (i,j)} }
 - 3 \bE{ \frac{S^{(3)}_i + S^{(3)}_j}{N (i,j)} }.
 \label{eq:Ex_mJCA_finalExpression1}
\end{equation}
We obtain the result by considering the different scaling regimes separately: The proof of Theorem~\ref{thm:MainTheorem} requires Lemmas~\ref{lemma:ProbabilisticResult} through 
\ref{lemma:Variance_Result} which are stated and proved in Appendix~\ref{sec:Lemma_proofs}.
\vspace{2mm}
\noindent
(a) First, consider the regime $p_n \to p$. From Lemma \ref{lemma:S1-Result} it is clear that all separate nodes are in sets $\cS ^{(1)} _{n,\ell}$ for $\ell=i,j$. Using this fact and \eqref{eq:C_n_behavior}, we obtain the result shown in \eqref{eq:lemma_ER_p_fixed}. 

\vspace{2mm}
\noindent
(b) First, we consider the regime $n p_n \to 0$. Under this regime, degree of nodes goes to $0$ a.s. 
Therefore for edge $(i,j)$, there are no common nodes and the separate nodes are $\{i,j\}$ whp. 
Therefore $\bE{gJC(i,j)} \to 0$.

\vspace{2mm}
\noindent
(c) Next, we consider the case $n p_n ^2 \to \infty$ and $p_n \to 0$. Fix $0< \epsilon < 1$: Suppose 
$n p_n ^2 = \Theta( n ^\epsilon)$. Therefore $n^{-\frac{\epsilon+1}{2}} n p_n = \Theta(1)$. 
Multiplying numerator and denominator in $\bE{ \frac{S_i ^{(1)}+S_j ^{(1)}+2}{N(i,j)} }$ 
by $n ^{-\frac{\epsilon+1}{2}}$ allows us to apply Lemma \ref{lemma:ProbabilisticResult}.
Using Lemma \ref{lemma:Variance_Result}, we note that
\begin{align}
Var \left( n ^{-\frac{\epsilon+1}{2}} N _n(1,2)  \right) &= 
n ^ { - (\epsilon +1) } Var (N_n(1,2)
\nonumber \\
& \sim n ^ { - (\epsilon +1) } n p_n = \Theta ( n^{-\frac{\epsilon+1}{2}} )
\label{eq:Denominator_Normalized_behavior_range1}
\end{align}
Eq. \eqref{eq:Denominator_Normalized_behavior_range1} and Lemmas \ref{lemma:S1-Result} yields $\bE{gJC(i,j)} \to 0$.

\vspace{2mm}
\noindent
(d) We consider the scaling range where $n ^2 p_n ^3 \to \infty$ and $n p_n ^2 \to 0$. 
Fix $\epsilon > 0$: Suppose $n ^2 p_n ^3 = \Theta (n ^\epsilon)$. Also, the constraint
$n p_n ^2 \to 0$ forces the bound $0 <\epsilon < \frac{1}{2}$. Therefore, 
$n ^{ -\frac{\epsilon+1}{3} } n p_n = \Theta (1)$.
 Multiplying the numerator and denominator of the individual terms of \eqref{eq:Ex_mJCA_finalExpression1} by $ n ^{ -\frac{\epsilon+1}{3} } $, 
 and observing that $ Var \left( n ^{-\frac{\epsilon+1}{3}} N _n(1,2) \right) 
 \sim n ^{-\frac{\epsilon+1}{3}} $, we obtain $\bE{gJC(i,j)} \to -1$ 
 by applying Lemmas \ref{lemma:S2_S3_Result} and \ref{lemma:Variance_Result}.

\vspace{2mm}
\noindent
(e) Now, we consider the scaling range where $n p_n \to \infty$ and $n ^2 p_n ^3 \to 0$. 
Fix $\epsilon >0$ and let $n p_n = \Theta (n ^\epsilon)$. The constraint $n ^2 p_n ^3 \to 0$ 
forces the bound $0 <\epsilon <\frac{1}{3}$. 
Multiplying the numerator and denominator of the individual terms of \eqref{eq:Ex_mJCA_finalExpression1} by $ n ^{ -\epsilon } $, we obtain 
$\bE{gJC(i,j)} \to -2$ 
 by applying Lemmas \ref{lemma:S2_S3_Result} and \ref{lemma:Variance_Result}.
 
\myendpf

Theorems \ref{thm:MainTheorem-JC}-\ref{thm:MainTheorem} together suggest that gJC is a better approximation of OR curvature than the JC curvature. We see that as the scaling changes and the ER graph becomes more dense, the gJC curvature increases progressively. In fact, the scalings at which the asymptotic behavior changes, match for the OR and gJC curvatures. The behavior of JC, gJC and OR curvatures are tabulated in Table \ref{table:JC_mJC_OR_ER} for different regimes in ER graphs.
 
\begin{table}[h]
\centering
\begin{tabular}{l|l|l|l|}
\cline{2-4}                                                                 & JC                 & gJC             & OR \\ \hline
\multicolumn{1}{|l|}{$p$ constant}                               & $\frac{5p-4}{2-p}$ & $\frac{p}{2-p}$ & $p$  \\ \hline
\multicolumn{1}{|l|}{$n p_n \to 0$}                              & $-2$                 & $0$               & $0$  \\ \hline
\multicolumn{1}{|l|}{$n p_n \to \infty$ and $n ^2 p_n ^3 \to 0$} & $-2$                 & $-2$              & $-2$ \\ \hline
\multicolumn{1}{|l|}{$n ^2 p_n ^3 \to \infty$ and $n p_n ^2 \to 0$}   & $-2$                 & $-1$  & $-1$ \\ \hline
\multicolumn{1}{|l|}{$n  p_n ^2 \to \infty$}                     & $-2$                 & $0$               & $0$  \\ \hline
\end{tabular}
\caption{The asymptotic values for the three curvatures under different scalings for the ER graph
\label{table:JC_mJC_OR_ER}
}
\end{table}


\externaldocument{ms.tex}
\externaldocument{curvature_setting.tex}
\externaldocument{theoretical_results.tex}

\section{Computational Complexity}
\label{sec:ComputationalComplexity}

Here we analytically study the complexity of computing the Ollivier-Ricci (OR), Forman, Jaccard (JC) and generalized Jaccard (gJC) curvatures of a graph. Intuitively there is a clear hierarchy in the complexity of the above-mentioned curvatures. Consider computing them for a generic edge $(i,j)$: In Forman we are only considering the degree of $i$ and $j$ so the complexity is going to be $O(m)$ for a graph of $m$ edges; while for JC, we are looking for the number of common neighbors between $i$ and $j$. In gJC we look for the shortest path to get from any exclusive neighbor of $i$ to any exclusive neighbor of $j$. All of these shortest paths could be to the same neighbor of $j$. Finally, in OR these shortest paths must represent a perfect fractional matching in the sense that one neighbor of y cannot be the target of too many neighbors of x. This is obviously a harder task.

For ease of computation we will first assume that the graph is d-regular and the graph is stored as sorted adjacency lists. In Appendix~\ref{sec:General_complexity}, we extend our analysis to address general graphs.

\subsection{Jaccard curvatures}

In Section~\ref{subsec:JCcurvature}, we divided $i,j$'s neighbor nodes into separate subsets:
\[
\cS_i^{(r)}  = \{ k \in \cN_i(i,j) \ | \ \min _{l \in \cN_j(i,j) } d(k,l) = r \}.
\]
\[
\cS_j^{(r)}  = \{ k \in \cN_j(i,j) \ | \ \min _{l \in \cN_i(i,j) } d(k,l) = r \}.
\]
for $r=0,1,2,3$. Let $\cS^{(r)}=\cS_i^{(r)} \cup \cS_j^{(r)}$, 
with $S^{(r)} = | \cS ^{(r)} |$, 
then   
\[
\cN(i,j) = \cS^{(0)}  \cup \cS^{(1)} \cup \cS^{(2)} \cup \cS^{(3)}   \cup \{i \cup j \}.
\] 



In order to compute JC and gJC, we need to count the size of $S^{(0)} , S^{(1)} , S^{(2)} , S^{(3)}$. In the following we will show how to compute these and their associated computational complexities. First, we note that computing the Forman curvature for all edges in the graph will be $O(m)$ or $O(nd)$, if the graph is d-regular.

\begin{lemma}
$S^{(0)}(i,j)$ can be computed with cost $O(d)$, and the total cost for the graph is $O(nd^2)$.
\label{lemma:S0}
\end{lemma}
\myproof
Let $\cN(i), \cN(j)$ be the sorted adjacency list of $i,j$, by merge sorting these two list; we get a new sorted list as $\cN(i,j)=\cN(i)\cup \cN(j)$. Then we have $\cC(i,j)=\cN(i)+\cN(j)-\cN(i,j)$ by inclusion–exclusion principle, therefore $S^{(0)}=2d-N(i,j)$. The merge-sort cost is $O(d)$. Since there are $m=O(nd)$ edges, the total cost of computing $S^{(0)}$ for all the edges in the graph will be $O(md)=O(nd^2)$
\myendpf

\begin{lemma}
$S^{(1)}(i,j)$ can be computed with cost $O(d^2)$, and the total cost for the graph is $O(nd^3)$.
\label{lemma:S1}
\end{lemma}
\myproof
Assume that we have $\cN(i)=(i_1,i_2,...i_d)$, $\cN(j)=(j,j_1,j_2,...j_d)$ as $i,j$'s sorted adjacency list. Let $\cN(i_s)$ be the sorted adjacency list of node $i_s$. As in Lemma~\ref{lemma:S0}, we merge-sort $\cN(i_s)$ and $\cN(j)$. If $\cN(i_s)\cap \cN(j)=\emptyset$, then $\min _{l \in \cN_i(j) } d(i_s,l) > 1$, and therefore we must have $i_s\notin S^{(1)}\cup S^{(0)}$; otherwise  if $\cN(i_s)\cap \cN(j)\not=\emptyset$,  then $i_s\in S^{(1)}\cup S^{(0)}$. We apply the same process to the $\cN(j)$ list. The total cost for computing $S^{(1)}(i,j)$ is $2d\times O(d)=O(d^2)$, and total cost for the graph is $O(md^2)=O(nd^3)$
\myendpf

\begin{lemma}
$S^{(2)}(i,j)$, $S^{(3)}(i,j)$ can be computed with cost of $O(d^3)$, the total cost for the graph is $O(nd^3)$.
\label{lemma:S2}
\end{lemma}
\myproof
Assume that we have sorted adjacency lists $\cN(i)=(i_1,i_2,...i_d)$,  $\cN(j)=(j,j_1,j_2,...j_d)$, $\cN(i_s)$, $\cN(j_t)$ defined as above. It is easy to see that for any two nodes $x,y$ if $N(x)\cap N(y)=\emptyset$, then $d(x,y)>2$. Therefore for each $i_s$, if for all $j_t\in \cN(j)$, we have $\cN(i_s)\cap \cN(j_t)=\emptyset$, then $\min _{l \in \cN(j) } d(i_s,l) > 2$, thus $i_s\in S^{(3)}$; if there exists $j_t\in \cN(j)$ such that $\cN(i_s)\cap \cN(j_t)\not=\emptyset$, then $\min _{l \in \cN_i(j) } d(i_s,l) \le 2$. With the help of $S^{(0)}(i,j),S^{(1)}(i,j)$ we determine if $i_s\in S^{(2)}(i,j)$. Therefore the total cost for confirming $i_s\in S^{(2)}(i,j)$ or $i_s \in S^{(3)}(i,j)$ is $O(d^2)$ and total cost for confirming all $\cN(i), \cN(j)$ is $2d\times O(d^2)=O(d^3)$.

The naive way of computing $S^{(2)}, S^{(3)}$ for the entire graph would be to apply this to each edge, with the resulting cost of $O(md^3)=O(nd^4)$. However we can save when computing $S^{(2)}(i,j), S^{(3)}(i,j)$ for all the edges in the graph. As a preprocessing step, we apply BFS on each nodes for depth at most $2$, Then we know the distance between all pairs of nodes~(distance could be $0,1,2,>2$).  By using this lookup table, we can determine if $d(i_s,j_t)>2$ with cost of $O(1)$ instead of $O(d)$ by applying merge-sort to $N(i_s)$ and $N(j_t)$. Therefore the total cost for determining $S^{(2)}(i,j), S^{(3)}(i,j)$ for all the edges could be reduced to $O(md^2)=O(nd^3)$. The cost for preprocessing of BFS is $O(nd^2)$ which is being dominated. So the total cost is $O(nd^3)+O(nd^2)=O(nd^3)$.

\myendpf

 .
\begin{theorem}
The cost for computing JC for the entire graph is $O(nd^2)$.
\label{thm:JC}
\end{theorem}

\myproof
Note that $\cS^{(0)}=\cC(i,j)$ and $2S^{(0)}+S^{(1)}+S^{(2)}+S^{(3)}=2d$. We also have  $\cS=\cS^{(1)}\cup\cS^{(2)}\cup\cS^{(3)}$. Therefore $S(i,j)=2d-2S^{(0)}$. Hence for JC we only need to compute $S^{(0)}$.

From Lemma~\ref{lemma:S0}, $S^{(0)}(i,j)$ can be computed with cost of $O(d)$, and $JC(i,j)$ with a cost of $O(d)$. The total cost in computing for the entire graph is $m\times O(d)=O(nd^2)$.
\myendpf

\begin{theorem}
The cost for computing gJC for the entire graph is $O(nd^3)$.
\label{thm:mJC}
\end{theorem}

\myproof
For gJC, we need to calculate all of $S^{(0)},S^{(1)},S^{(2)},S^{(3)}$ . From Lemmas~\ref{lemma:S0}-\ref{lemma:S2}, the total cost will be $O(nd^3)$.
\myendpf

\subsection{Ollivier-Ricci curvature}
The computation of the Olliver-Ricci curvature of an edge $(i,j)$ is an optimization problem. It is usually solved by an LP solver which is not guaranteed  to be in polynomial time. Here we translate the OR curvature problem into a min-cost max flow problem as follows:

First we create a complete bipartite graph $G=(I\cup J, E)$, where $I=\cN(i)$ and $J=\cN(j)$, with $\cN_i$ being the neighbor set of $i$. Cost of edge $(i_s, j_t)$, $c_{s,t}$, is set to $d(i_s, j_t)$ in the original graph, and the capacity of each edge set to infinity. Next, we add a source node $x$ and sink node $y$; source node $x$ connects to all the nodes in $I$ with cost of edge $(x,i_s)$,  $c_{x,s}$, set to $1$ and capacity set to the mass $m_{i_s}$ distributed on the original graph; the sink node $y$ connects to all the nodes in $J$ with cost of edge $(y,j_t)$, $c_{y,t}$, set to $1$ and capacity  set to the mass $m_{j_t}$. The goal is to minimize the total cost along the edge with maximum possible flow $f$, where the cost is defined as
$C=\sum_{e\in E} c_e\times f(e)$.


We can solve the min-cost max flow problem using the network simplex algorithm~\cite{tarjan1997dynamic}. 
\begin{theorem}
The cost for computing OR curvatures for the entire graph is $O(nd^4\log^2 d)$
\end{theorem}

\myproof
In Tarjan's paper~\cite{tarjan1997dynamic}, it was shown that the network simplex algorithm has complexity $O(mn \log n \log C)$, where $m,n$ are the number of nodes and edges in the graph respectively, and $C$ is the maximum edge cost. In the present setting, $m=d^2+2d, n=2d+2, C=1/\frac{1}{d}=d$, so computing OR curvature for an edge will have complexity $O(d^3\log^2d)$, with the total complexity for the entire graph being $O(nd^4\log^2 d)$.

Note that the algorithm is for a general graph, but for ease of computation complexity we still use d-regular setting here. Results on the computational complexity for general graphs are provided in Appendix~\ref{sec:General_complexity}.
\myendpf





\begin{table}[h!]
\centering
\begin{tabular}{|c|c|c|c|c|}
\hline
& Forman & JC & gJC & OR \\
\hline
Complexity & $O(nd)$ & $O(nd^2)$ & $O(nd^3)$ & $O(n^4\log^2 n)$ \\
\hline
\end{tabular}
\label{complexity hierarchy}
\caption{The complexity hierarchy for Forman, JC, gJC and OR curvatures}
\end{table}

\section{Experimental Results}
\label{sec:experimental_results}

\subsection{Network models}
We consider different network models to investigate the relationship between Jaccard and Forman curvatures in relation to Ollivier-Ricci curvature. Specifically, we explore network models for which Jaccard curvature or Forman curvature gives a reasonably good approximation to Ollivier-Ricci curvature. This would inform types of real-world networks for which Jaccard curvatures could be used to approximate Ollivier-Ricci curvature. The network models being considered are described below. 

\textbf{Erd{\"o}s-R{\'e}nyi (ER) model}: The ER model introduced by Erd{\"o}s and R{\'e}nyi~\cite{erdos1960evolution} has long been considered to be a suitable mathematical model for networks because of the simplicity and mathematical tractability in analyzing their properties. $ER(n,p)$ is a network on $n$ nodes that connects every pair of nodes with probability $p$ independent across node pairs. We fix $n=100$ and vary $p$ from $0.05$ to $0.9$, to study the behavior of the Jaccard and Forman curvatures vis-a-vis Ollivier-Ricci curvature.

\begin{figure}[h]
\centering
\begin{subfigure}{.55\textwidth}
  \centering
  \includegraphics[width=.8\linewidth]{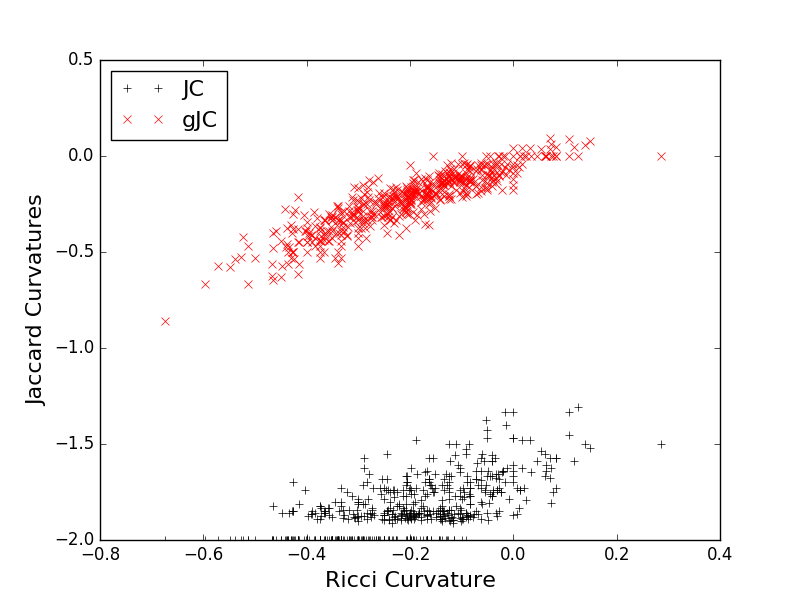}
  \caption{ER(100,0.1)}
  \label{fig:ER_100_01}
\end{subfigure}%
\begin{subfigure}{.55\textwidth}
  \centering
  \includegraphics[width=.8\linewidth]{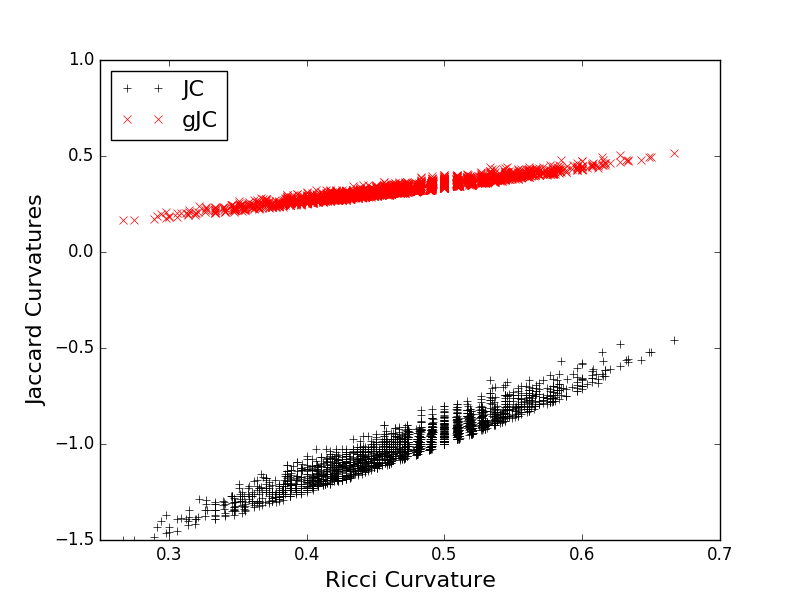}
  \caption{ER(100,0.5)}
  \label{fig:sub2}
\end{subfigure}
\begin{subfigure}{.55\textwidth}
  \centering
  \includegraphics[width=.8\linewidth]{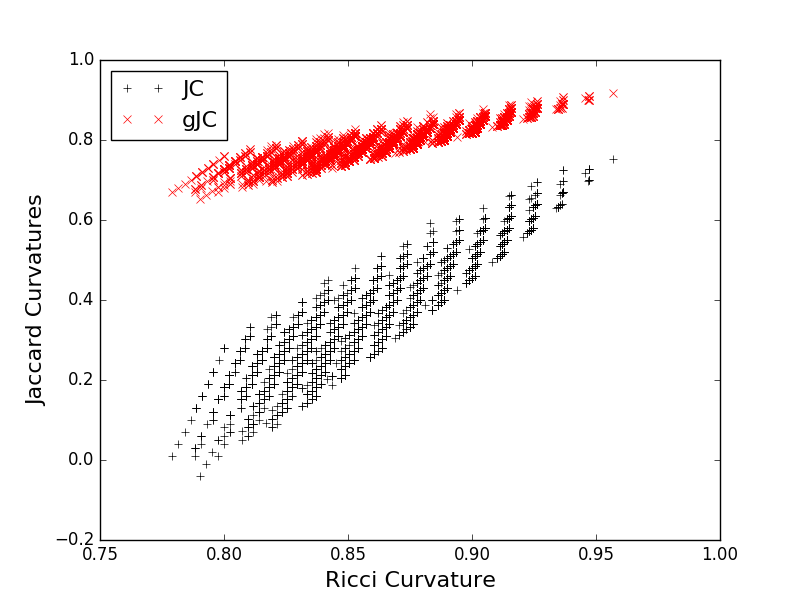}
  \caption{ER(100,0.9)}
  \label{fig:ER_100_09}
\end{subfigure}%
\caption{Scatter plots of Ollivier-Ricci and Jaccard curvatures for Erd{\"o}s-R{\'e}nyi graphs}
\label{fig:ER_scatterplots}
\end{figure}

Table \ref{table:ER_results} shows the average curvatures of different ER graphs as the probability of connection $p$ is varied. We observe that for {\textit small $p$}, the average OR curvature is negative, and as $p$ increases the  average OR curvature increases as well. This is because as $p$ increases, the density of the graph increases leading to more edges with positive curvature. For $p$ small, the ER graph is more disconnected and tree-like, leading to negative average OR curvature. This behavior is well replicated by the Jaccard curvatures. In particular, the average gJC curvature closely tracks the average OR curvature as $p$ is varied. On the other hand, the average Forman curvature decreases as $p$ is increased for the simple reason that the average degree of nodes increase. We also observe that gJC correlates the best with OR curvature, and this difference is more pronounced for ER graphs with low $p$ value. The advantage of gJC with respect to JC is somewhat lost as $p$ is increased beyond $0.2$. Furthermore, the correlation between the Jaccard curvatures and OR curvature improves as $p$ is increased, but deteriorates slightly as $p$ is increased beyond $0.5$. Scatter plots shown in Figure~\ref{fig:ER_scatterplots} provides a visual representation of the correlation between OR and Jaccard curvatures. It is clear visually that the spread in the scatter plot is the least for $p=0.5$, which is also supported by the correlation coefficients shown in the table.

\begin{table}[h]
\centering
\begin{tabular}{|l|l|l|l|l|l|l|l|l|l|l|}
\hline
\multirow{2}{*}{Graph} & \multirow{2}{*}{$\overline{OR}$} & \multirow{2}{*}{$\overline{JC}$} & \multirow{2}{*}{$\overline{gJC}$} & \multirow{2}{*}{$\overline{F}$} & \multicolumn{2}{l|}{(OR,JC)} & \multicolumn{2}{l|}{(OR,gJC)} & \multicolumn{2}{l|}{(OR,F)} \\ \cline{6-11} 
                       &                     &                     &                      &                    & $r_p$           & $\tau$           & $r_p$            & $\tau$           & $r_p$           & $\tau$          \\ \hline \hline
ER(100,0.05)	&  -0.59  &	-1.95  &	-0.86  &	-7.8  &	0.4  &	0.28  &	0.77  &	0.55  &	0.35  &	0.25
            \\ \hline
  ER(100,0.1)  &	-0.20  &	-1.83  &	-0.23  &	-19.3  &	0.64  &	0.45  &	0.90  &	0.73  &	-0.31  &	-0.21
            \\ \hline
  ER(100,0.2)  &	0.15  &	-0.77  &	0.09  &	-37.4  &	0.89  &	0.76  &	0.94  &	0.80  &	-0.46  &	-0.3
           \\ \hline
  ER(100,0.3)  &	0.26  &	-1.48  &	0.17  &	-57  &	0.97  &	0.86  &	0.97  &	0.86  &	-0.53  &	-0.38
            \\ \hline
  ER(100,0.4)  &	0.35  &	-1.3  &	0.23   &	-74  &	0.97  &	0.86  &	0.97   &	0.86  &	-0.44  &	-0.28
            \\ \hline
  ER(100,0.5)  &	0.47  &	-1.01  &	0.35  &	-97.12  &	0.96 &	0.83 &	0.96  &	0.83  &	-0.54  &	-0.37
               \\ \hline
  ER(100,0.6)  &	0.55  &	-0.77  &	0.41  &	-113  &	0.93  &	0.77  &	0.93   &	0.77  &	-0.58   &	-0.4
            \\ \hline
 ER(100,0.7)  &	0.67   &	-0.39  &	0.54   &	-137  &	0.92 &	0.75  &	0.92   &	0.75  &	-0.64  &	-0.45
             \\ \hline
 ER(100,0.8)  &	0.77   &	-0.06 &	0.65   &	-154.03  &	0.91  &	0.74 &	0.91  &	0.73  &	-0.7  &	-0.5
             \\ \hline
 ER(100,0.9)  &	0.87  &	0.38  &	0.8   &	-173 &	0.91   &	0.77  &	0.91  &	0.77  &	-0.83  &	-0.68
             \\ \hline
\end{tabular}
\caption{Average curvatures shown for different Erdos Renyi (ER) graphs [ER(n,p) - number of nodes being n, and probability of connection being p]. The Pearson correlation coefficient $r_p$ and Kendall's $\tau$ coefficient between the OR and Jaccard curvatures, and the OR and Forman curvature are tabulated.
\label{table:ER_results}
}
\end{table}

\textbf{Barab{\'a}si-Albert (BA) model}: The BA model introduced by Barab{\'a}si and Albert~\cite{barabasi1999emergence} became popular and widely used because it was the first network model to explain power-law behavior through the preferential attachment or the rich-get-richer phenomena. $BA(n,m)$ is a network growth model, which starts with $m$ nodes and new nodes connect preferentially to $m$ existing nodes with probability proportional to their degree. We study how the curvature metrics behave as $n$ and $m$ are varied. So first we fix $n=100$ and consider $m=1,2,5$, and then we study networks with a larger value of $n=500$.

\begin{figure}[h]
\centering
\begin{subfigure}{.55\textwidth}
  \centering
  \includegraphics[width=.8\linewidth]{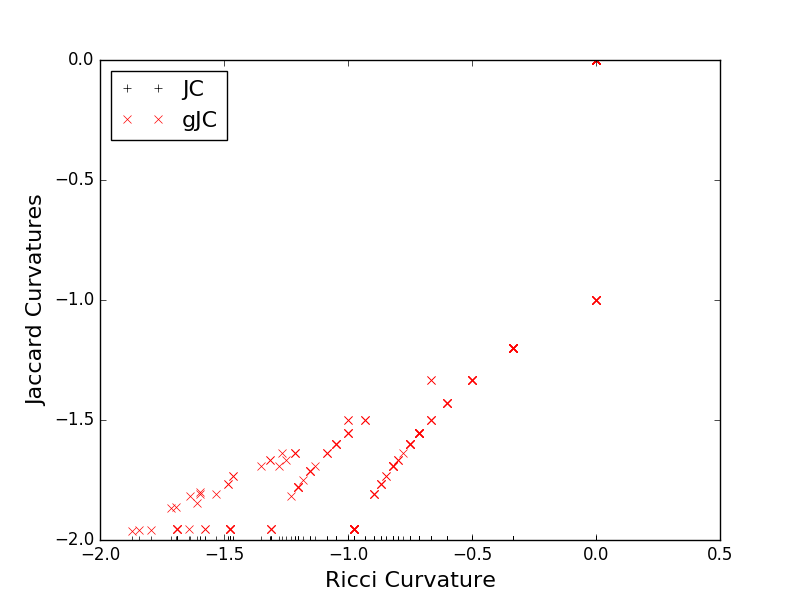}
  \caption{BA(500,1)}
  \label{fig:BA_500_1}
\end{subfigure}%
\begin{subfigure}{.55\textwidth}
  \centering
  \includegraphics[width=.8\linewidth]{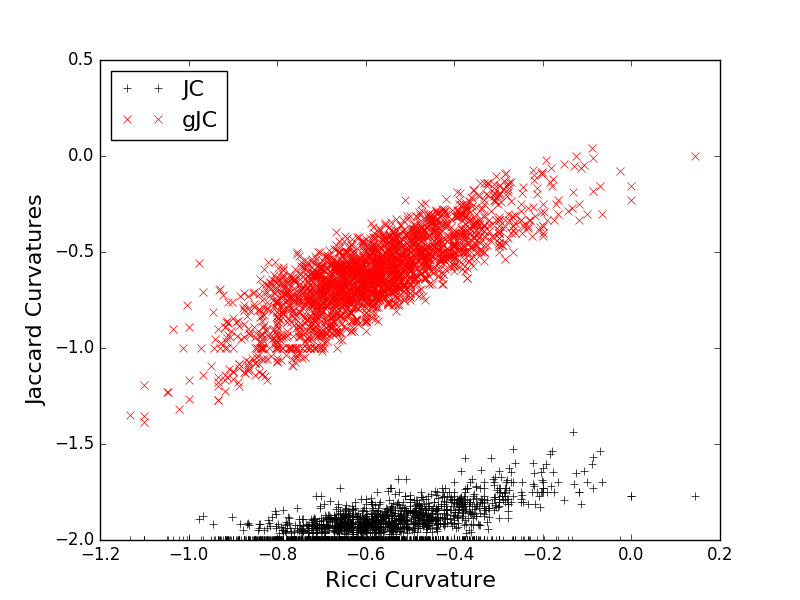}
  \caption{BA(500,5)}
  \label{fig:BA_500_5}
\end{subfigure}
\caption{Scatter plots of Ollivier-Ricci and Jaccard curvatures for Barab{\'a}si-Albert graphs}
\label{fig:BA_scatterplots}
\end{figure}

In Table \ref{table:BA_results} we observe that gJC correlates the best with OR curvature for BA(100,1), a dramatic improvement over JC which has a single value of $-2$ for all the edges. Forman curvature correlates better than JC for BA(100,1), suggesting that it captures some properties of OR curvature for sparse tree-like graphs. Because there are no triangles in BA(100,1), JC simply cannot capture any information. On the other hard, gJC performs very well in approximating OR, because all mass transport paths need to pass through the edge being considered, leading to most mass transports requiring 3 hop paths, except the endpoints of the edge itself. Increasing $m$ from 1 to 2 and keeping number of nodes fixed at $100$, results in slightly deterioration of the correlation between gJC and OR curvature because now there could be shorter than 3 hop paths that need to be accounted for. However, increasing $m$ further leads to improvement, similar to what was observed in ER graphs for moderate $p$ values. Increasing number of nodes $n$ to $500$ and keeping $m$ fixed, decreases the correlation of Jaccard curvatures slightly, probably because that makes the graph more tree-like with larger hubs. Scatter plots in Figure~\ref{fig:BA_scatterplots} show that the spread is higher in BA graph compared to ER graphs. 

\begin{table}[h]
\centering
\begin{tabular}{|l|l|l|l|l|l|l|l|l|l|l|}
\hline
\multirow{2}{*}{Graph} & \multirow{2}{*}{$\overline{OR}$} & \multirow{2}{*}{$\overline{JC}$} & \multirow{2}{*}{$\overline{gJC}$} & \multirow{2}{*}{$\overline{F}$} & \multicolumn{2}{l|}{(OR,JC)} & \multicolumn{2}{l|}{(OR,gJC)} & \multicolumn{2}{l|}{(OR,F)} \\ \cline{6-11} 
                       &                     &                     &                      &                    & $r_p$           & $\tau$           & $r_p$            & $\tau$           & $r_p$           & $\tau$          \\ \hline \hline
BA(100,1)  &	-0.31  &	-2  & 	-0.54  &	-3.84  &	N/A  &	N/A &	0.92  &	0.94  &	0.65 &	0.5
            \\ \hline
BA(100,2) &	-0.45  &	-1.9  &	-0.85  &	-10.26  &	0.47 &	0.24  &	0.62  &	0.45  &	0.6   &	0.5
            \\ \hline
BA(100,5)  &	-0.16 &	-1.77  &	-0.19  &	-25  &	0.73  &	0.54  &	0.86  &	0.66  &	-0.16  &	-0.07
           \\ \hline
BA(500,1)  &	-0.31  &	-2  &	-0.55  &	-16.8  &	N/A  &	N/A  &	0.93  &	0.72  &	0.33  &	0.37
		\\ \hline
BA(500,2)  &	-0.79  &	-1.98  &	-1.32  &	-11.7  &	0.12  &	-0.05  &	0.52  &	0.4  &	0.64  &	0.58
    \\ \hline
BA(500,5)  &	-0.58  &	-1.94  &	-0.62  &	-32 &	0.52  &	0.34  &	0.81  &	0.6  &	-0.08  &	0.03
            \\ \hline
\end{tabular}
\caption{Average curvatures shown for different Barabasi-Albert graphs [BA(n,m) - m being the number of connections formed by a new node]. The Pearson correlation coefficient $r_p$ and Kendall's $\tau$ coefficient between the OR and Jaccard curvatures, and the OR and Forman curvature are tabulated.
\label{table:BA_results}
}
\end{table}

\textbf{Watts-Strogatz (WS) model}: The WS model introduced by Watts and Strogatz~\cite{watts1998collective} produces graphs with small-world properties, exhibiting short average path lengths and high clustering coefficent.   $WS(n,k,p)$ is a network model on $n$ nodes which first constructs a ring among adjacent nodes such that each node is connected to $k$ closest neighbors. Then every edge is randomly rewired with probability $p$ by keeping one endpoint fixed and choosing the other endpoint uniformly at random. We show curvature results for $p=0.1$ and $k=4$, and vary $n=100,200,500$. Furthermore, we show results for $n=500$,$k=4$ and varying $p=0.1,0.3,0.5,0.7,0.9,0.99$. Thus, we analyze the results as the number of nodes and the rewiring probability varies. 

\begin{figure}[h]
\centering
\begin{subfigure}{.55\textwidth}
  \centering
  \includegraphics[width=.8\linewidth]{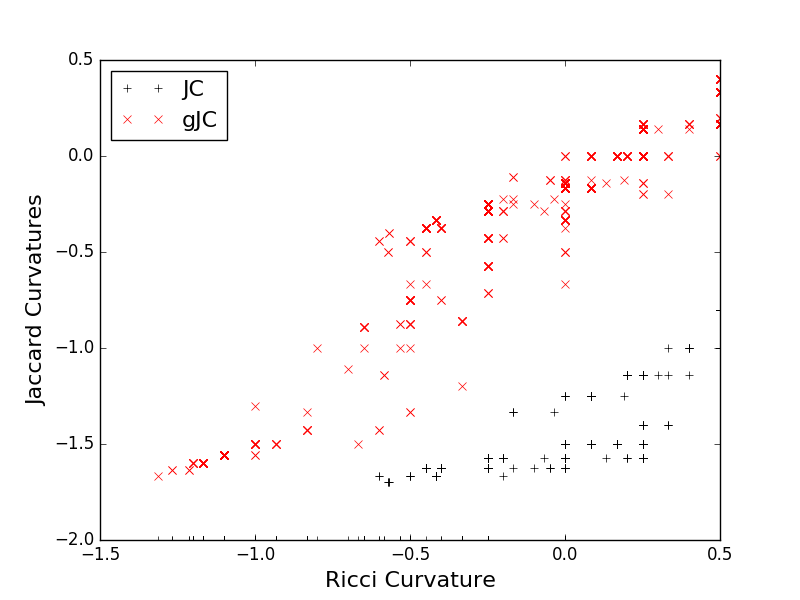}
  \caption{WS(500,0.1,4)}
  \label{fig:WS_500_01}
\end{subfigure}%
\begin{subfigure}{.55\textwidth}
  \centering
  \includegraphics[width=.8\linewidth]{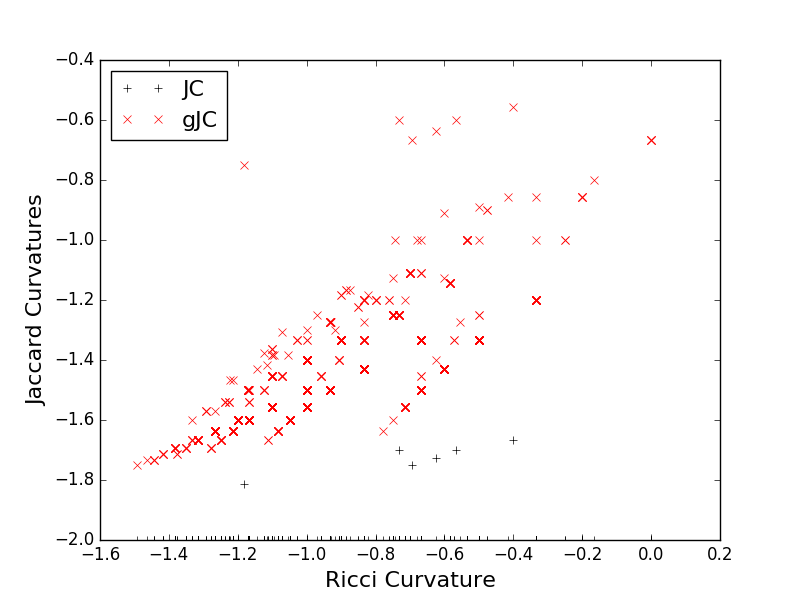}
  \caption{WS(500,0.99,4)}
  \label{fig:WS_500_099}
\end{subfigure}
\caption{Scatter Plots of Ollivier-Ricci and Jaccard curvatures for Watts Strogatz graphs}
\label{fig:WS_scatterplots}
\end{figure}

Table \ref{table:WS_results} shows the average curvatures and correlation results for different WS networks. We observe that keeping the parameters $p$ and $k$ fixed, increasing the number of nodes $n$ does not change the curvatures much. Increasing the probability of rewiring makes the curvature more negative. This behavior is shown by OR, JC, gJC and the Forman curvatures. Ollivier-Ricci and Jaccard curvatures show this behavior because increasing the rewiring probability, increases the number of shortcuts in the network, thus making the average curvature more negative. Increasing the rewiring probability leads to a steady decrease in correlation between the Jaccard and the OR curvature, with Forman curvature showing the opposite behavior. This leads us to believe that the Jaccard curvatures are a better fit for the OR curvature for more positively curved graphs. Nevertheless, for $WS(500,4,0.99)$ where the average OR curvature is $-0.93$, the correlation coefficients between gJC and OR are still pretty high. 
Figure~\ref{fig:WS_scatterplots} shows how the spread in the scatter plots increases as the probability of rewiring is increased from $0.1$ to $0.99$. 
\begin{table}[h]
\centering
\begin{tabular}{|l|l|l|l|l|l|l|l|l|l|l|}
\hline
\multirow{2}{*}{Graph} & \multirow{2}{*}{$\overline{OR}$} & \multirow{2}{*}{$\overline{JC}$} & \multirow{2}{*}{$\overline{gJC}$} & \multirow{2}{*}{$\overline{F}$} & \multicolumn{2}{l|}{(OR,JC)} & \multicolumn{2}{l|}{(OR,gJC)} & \multicolumn{2}{l|}{(OR,F)} \\ \cline{6-11} 
                       &                     &                     &                      &                    & $r_p$           & $\tau$           & $r_p$            & $\tau$           & $r_p$           & $\tau$          \\ \hline \hline
WS(100,4,0.1)  &	-0.04  &	-1.46  &	-0.22  &	-4.17  &	0.89  &	0.89  &	0.96 &	0.9 &	0.33  &	0.25
            \\ \hline
WS(200,4,0.1)  &	-0.08 &	-1.51  &	-0.28  &	-4.2 &	0.88  &	0.86  &	0.96  &	0.9  &	0.37  &	0.26
            \\ \hline
WS(500,4,0.02)  &	0.17  &	-1.33  &	0.02  &	-4.04 &	0.93  &	0.95  &	0.98  &	0.95  &	0.29  &	0.21
           \\ \hline
WS(500,4,0.05) &	0.06  &	-1.41   &	-0.11  &	-4.1  &	0.9  &	0.91  &	0.96  &	0.92  &	0.29 &	0.23
            \\ \hline
WS(500,4,0.1) &	-0.05   &	-1.47  &	-0.24 &	-4.2  &	0.89 &	0.87  &	0.96 &	0.9 &	0.42  &	0.31
            \\ \hline
WS(500,4,0.2) &	-0.31  &	-1.65  &	-0.57   &	-4.36  &	0.86 &	0.79  &	0.95  &	0.87   &	0.39 &	0.3
   \\ \hline
WS(500,4,0.3)  &	-0.54  &	-1.78  &	-0.88  &	-4.5  &	0.85 &	0.73 &	0.94  &	0.85  &	0.42  &	0.34
  \\ \hline
WS(500,4,0.5)  &	-0.77  &	-1.91   &	-1.2  &	-4.7  &	0.77  &	0.55  &	0.89  &	0.83 &	0.49 &	0.45
\\ \hline
WS(500,4,0.7)  &	-0.9  &	-1.98  &	-1.4  &	-4.8  &	0.52  &	0.3 &	0.8  &	0.76  &	0.66  &	0.61
\\ \hline
WS(500,4,0.9)  &	-0.94  &	-1.99  &	-1.45  &	-4.8 &	0.35 &	0.19  &	0.78  &	0.78  &	0.76  &	0.69
\\ \hline
WS(500,4,0.99)  &	-0.93  &	-2  &	-1.45  &	-5  &	0.08  &	0.06  &	0.78   &	0.68  &	0.76  &	0.65
\\ \hline 
\end{tabular}
\caption{Average curvatures shown for different Watts Strogatz graph [WS(n,k,p) - each node connects to k nearest neighbors and p is the probability of rewiring an edge]. The Pearson correlation coefficient $r_p$ and Kendall's $\tau$ coefficient between the OR and Jaccard curvatures, and the OR and Forman curvature are tabulated.
\label{table:WS_results}
}
\end{table}

\textbf{Random geometric graph (RGG) model}: 
Random geometric graphs~\cite{penrose2003random} are spatial networks that have found application in the modeling of ad hoc mobile networks~\cite{nekovee2007worm}.
$RGG(n,r)$ is a network model where all nodes are distributed uniformly on a metric space, e.g., a unit square, and connections between nodes are formed only if the pairwise Euclidean distance is less than a certain radius $r$, with $0<r<1$. We study the curvature results for fixed $n=500$ and varying radius $r$.

\begin{figure}[h]
\centering
\begin{subfigure}{.55\textwidth}
  \centering
  \includegraphics[width=.8\linewidth]{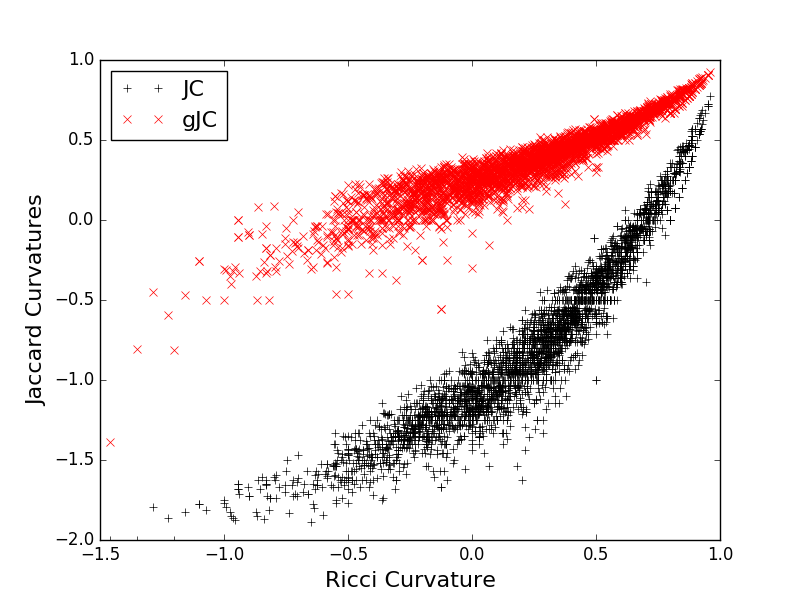}
  \caption{RGG(500,0.1)}
  \label{fig:RGG_500_01}
\end{subfigure}%
\begin{subfigure}{.55\textwidth}
  \centering
  \includegraphics[width=.8\linewidth]{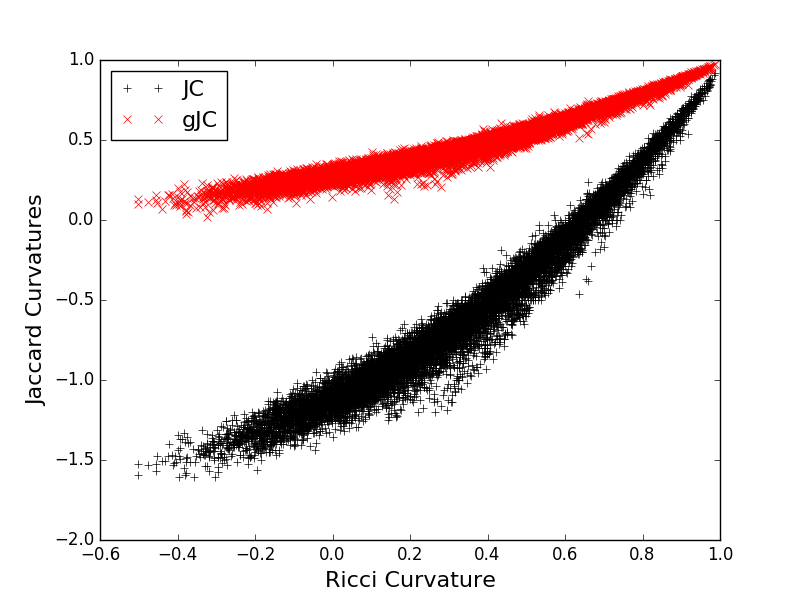}
  \caption{RGG(500,0.2)}
  \label{fig:RGG_500_02}
\end{subfigure}
\caption{Scatter Plots of Ollivier-Ricci and Jaccard curvatures for random geometric graphs}
\label{fig:RGG_scatterplots}
\end{figure}

From Table~\ref{table:RGG_results}, we observe that increasing the radius $r$ increases the OR curvature slightly. This is because clustering of the network increases with increasing $r$, thus leading to an increase in the OR curvature. Both the Jaccard curvatures exhibit this behavior too, although the mean gJC curvature values are much closer to the mean OR curvature values. On the other hand, the Forman curvature becomes more negative as $r$ increases, because of an increase in the average degree of nodes. Furthermore, the correlation coefficients of the Jaccard curvatures increase as $r$ increases. This observation agrees with the previously mentioned hypothesis that the Jaccard curvature approximates OR curvature better for positively curved graphs. Figure~\ref{fig:RGG_scatterplots} visually shows how the fit of the Jaccard curvature improves as the radius of connectivity $r$ is increased.

\begin{table}[h]
\centering
\begin{tabular}{|l|l|l|l|l|l|l|l|l|l|l|}
\hline
\multirow{2}{*}{Graph} & \multirow{2}{*}{$\overline{OR}$} & \multirow{2}{*}{$\overline{JC}$} & \multirow{2}{*}{$\overline{gJC}$} & \multirow{2}{*}{$\overline{F}$} & \multicolumn{2}{l|}{(OR,JC)} & \multicolumn{2}{l|}{(OR,gJC)} & \multicolumn{2}{l|}{(OR,F)} \\ \cline{6-11} 
                       &                     &                     &                      &                    & $r_p$           & $\tau$           & $r_p$            & $\tau$           & $r_p$           & $\tau$          \\ \hline \hline
RGG(500,0.05) &	0.22  &	-1.13  &	0.05  &	-4.89  &	0.86  &	0.83  &	0.88  &	0.79  &	-0.17  &	-0.15
            \\ \hline
RGG(500,0.1)  &	0.23  &	-0.75 &	0.37  &	-28  &	0.94  &	0.86  &	0.94  &	0.82  &	-0.22  &	-0.12
            \\ \hline
RGG(500,0.15)  &	0.28  &	-0.64  &	0.44  &	-64.4  &	0.96  &	0.87  &	0.96  &	0.86 &	0.04  &	0.04
           \\ \hline
RGG(500,0.2)  & 0.32  &	-0.61  &	0.46 &	-112  &	0.97  &	0.89  &	0.97  &	0.88  &	0.16  &	0.09
 \\ \hline
\end{tabular}
\caption{Average curvatures shown for different Random Geometric Graphs [RGG(n,r) - each node connects to nodes within a distance of r]. The Pearson correlation coefficient $r_p$ and Kendall's $\tau$ coefficient between the OR and Jaccard curvatures, and the OR and Forman curvature are tabulated.
\label{table:RGG_results}
}
\end{table}

\subsection{Real-world networks}
We consider several real-world networks in this subsection. The Gnutella network has been obtained from the Stanford large network dataset collection~\cite{snapnets}, while the rest of the networks were obtained from the Koblenz Network Collection~\cite{konect}. A brief description of the datasets is provided below. Their network properties are displayed in Table~\ref{table:network_properties}.
\begin{table}[h]
\centering
\begin{tabular}{|l|l|l|l|l|l|l|l|l|}
\hline
Dataset        & n     & m     & $d_{max}$ & $d_{avg}$ & Diameter & \begin{tabular}[c]{@{}l@{}}Mean shortest\\ path length\end{tabular} & \begin{tabular}[c]{@{}l@{}}Clustering \\ coefficient\end{tabular} & Assortativity \\ \hline \hline
US power grid  & 4941  & 6594  & 19         & 2.67        & 46       & 20                                                                  & 0.1                                                               & 0.003         \\ \hline
EuroRoad       & 1174  & 1417  & 10         & 2.41        & 62       & 19                                                                  & 0.03                                                              & 0.13          \\ \hline
PGP network    & 10680 & 24316 & 205        & 4.55        & 24       & 7.65                                                                & 0.38                                                              & 0.24          \\ \hline
p2p-Gnutella       & 6301  & 20777 & 97       & 6.59        & 9        & 4.64                                                                & 0.01                                                              & 0.03          \\ \hline
Email network  & 1133  & 5451  & 71         & 9.62        & 8        & 3.65                                                                & 0.17                                                              & 0.08          \\ \hline
Hamsterster    & 1858  & 12534 & 272        & 13.5        & 14       & 3.4                                                                 & 0.09                                                              & -0.08         \\ \hline
Human protein  & 3133  & 6726  & 129        & 4.29        & 13       & 4.80                                                                & 0.04                                                              & -0.13         \\ \hline
Jazz musicians & 198   & 2742  & 100        & 27.69       & 6        & 2.21                                                                & 0.52                                                              & 0.02          \\ \hline
\end{tabular}
\caption{Network properties of real-world networks being considered. The number of nodes $n$, number of edges $m$, maximum degree $d_{max}$, average degree $d_{avg}$, and other well-known network properties are reported.
\label{table:network_properties}
}
\end{table}
\subsubsection{Description of networks:}

\textbf{Infrastructure Networks:} We study the US Power Grid network~\cite{watts1998collective} which contains information about the power grid of the Western states of the United States of America. A node in the network is either a generator, a transformer or a power substation, while edges represent high-voltage power supply lines between nodes. This network was studied in \cite{watts1998collective} as an example of a real network with the small-world property.  Another infrastructure network we consider is the Euroroad network~\cite{vsubelj2011robust}, a road network located mostly in Europe. The network is undirected, with nodes representing cities and an edge between two nodes denotes a physical road between them. This network was observed to be neither scale-free nor small-world, and particularly difficult to partition with standard community detection algorithms.

\noindent
\textbf{Online social and communication networks:} 
We study a social network formed by people that shares confidential information using the Pretty Good Privacy (PGP) encryption algorithm, also called the PGP web of trust~\cite{boguna2004models}. The degree distribution of the network exhibits a power law decay with a particular exponent for small degrees, and a crossover towards another power-law with a higher exponent for large degree values. This suggests that unlike many technological networks, the PGP is not a scale-free network but exhibits bounded degree distribution. Furthermore, the network shows a large clustering coefficient.
We also consider the Gnutella network~\cite{ripeanu2002mapping}, a peer-to-peer architecture, where nodes represent Gnutella hosts and the edges represent connections between them. It is not a pure power-law network and preserves good fault tolerance characteristics, while being less dependent than a pure power-law network on highly connected nodes. The email communication network~\cite{guimera2003self} at the University Rovira i Virgili in Tarragona is also studied. Here, the nodes of the network are users, and each edge represents that at least one mail was exchanged between the concerned nodes.
This network was studied as an example of a self-organized complex system~\cite{guimera2003self}. 
 
\noindent
\textbf{Other miscellaneous networks:}
We considered a biological network that was an initial version of a systematic mapping of protein-protein interactions in humans~\cite{rual2005towards}. We also considered a collaboration network between Jazz musicians, with each node being a Jazz musician and an edge denoting the two musicians playing together in a band~\cite{gleiser2003community}.

\subsubsection{Discussion of results}

Table~\ref{table:RealWorldNetworks_results} shows the mean curvature values and correlation coefficients between OR and the other curvature metric for different real-world networks. Firstly, we observe that gJC tracks OR the closest in terms of the mean curvature. Since the range of Forman is not bounded, we see large negative average curvatures for many networks, unlike Jaccard and OR curvatures which are inherently bounded.  
Furthermore, the gJC curvature correlates strongly with OR compared to JC and Forman curvatures for almost every real network being considered. Even JC seems to outperform Forman curvature in correlating with OR curvature on PGP Network, p2p-Gnutella, Email network, Hamsterster friendship network and Human protein network, while Forman correlates stronger with OR solely on the EuroRoad network. 

Table~\ref{table:RealWorldNetworks_running_time} shows that the gJC implementation is several orders of magnitude faster than the OR implementation, while the JC implementation itself is faster compared to the gJC implementation. 
The Forman implementation is the fastest among all the curvature metrics being considered, thus agreeing with the theoretical analysis in Section~\ref{sec:ComputationalComplexity}. However, the results suggest that there is no clear scenario where the Forman is a good proxy for OR curvature, while JC could be a good proxy for OR for positively curved or more clustered networks. Although the gJC curvature is computationally costlier than JC and Forman, it correlates strongly with OR curvature for most of the real networks being considered across a wide range of network properties and types. 

\begin{table}[h]
\centering
\begin{tabular}{|l|l|l|l|l|l|l|l|l|l|l|}
\hline
\multirow{2}{*}{Graph} & \multirow{2}{*}{$\overline{OR}$} & \multirow{2}{*}{$\overline{JC}$} & \multirow{2}{*}{$\overline{gJC}$} & \multirow{2}{*}{$\overline{F}$} & \multicolumn{2}{l|}{(OR,JC)} & \multicolumn{2}{l|}{(OR,gJC)} & \multicolumn{2}{l|}{(OR,F)} \\ \cline{6-11} 
                       &                     &                     &                      &                    & $r_p$           & $\tau$           & $r_p$            & $\tau$           & $r_p$           & $\tau$          \\ \hline \hline
US power grid  &	-0.34  &	-1.89  &	-0.78  &	-3.7  &	0.4  &	0.23  &		0.8  &	0.69  &		0.48  &	0.41
            \\ \hline
EuroRoad  &	-0.33  &	-1.97  &	-0.97  &	-2	&		0.15  &	0.09  &		0.69  &	0.67 &		0.76  &	0.69
 \\ \hline
PGP network  &	-0.10  &	-1.36  &	-0.14  &	-33.76  &			0.73  &	0.53  &		0.85  &	0.74    &		0.13  &	0.08
            \\ \hline
p2p-Gnutella  &	-1.01  &	-1.98  &	-1.16  &	-31  &			0.23  &	0.27  &		0.86	 &  0.58	&	-0.32  &	0.08
           \\ \hline
Email network  &	-0.41  &	-1.72  &	-0.38  &	-33.37  &			0.73  &	0.56	&	0.81  &	0.69  &		0.15  &	0.11
\\ \hline
Hamsterster friendships & -0.34 & -1.87 & -0.19 & -86.7 & 0.41 & 0.23 & 0.58 & 0.42 &0.13 & 0.13 \\ 
\hline
Human protein  &	-0.62  &	-1.93  &	-0.79  &	-27   &			0.31  &	0.1	  &	0.78  &	0.6  &		0.35   &	0.34
   \\ \hline
Jazz musicians &	0.27  &	-0.92  &	0.32  &	-73.3    &			0.91  &	0.79  &		0.92  &	0.8   &		0.09 &	0.05
   \\ \hline
\end{tabular}
\caption{Average curvatures shown for different real-world networks. The Pearson correlation coefficient $r_p$ and Kendall's $\tau$ coefficient between the OR and Jaccard curvatures, and the OR and Forman curvature are tabulated.
\label{table:RealWorldNetworks_results}
}
\end{table}

\begin{table}[h!]
\centering
\begin{tabular}{|c|c|c|c|c|}
\hline
& OR(LP solver)&gJC & JC & Forman\\
\hline
\hline
US power grid  &  146.011s & 0.67s &  0.368s &  0.099s  \\
\hline
EuroRoad  & 9.515s &  0.432s & 0.227s &  0.023s \\
\hline
PGP network &  1052.614s &  5.258s &  2.624s  &  0.419s \\
\hline
p2p-Gnutella   & 219.943s &  5.66s  &  2.146s   &  0.331s\\
\hline
Email network &  57.279s &   1.543s &  0.718s &  0.071s\\
\hline
Hamsterster friendships  &  424.898s &   14.354s & 5.069s &  0.267s \\
\hline
Human protein &  78.64s &    1.312s &   0.951s &   0.084s\\
\hline
Jazz musicians  & 72.137s &   1.539s &  0.957s &  0.092s \\
\hline
\end{tabular}
\caption{Running times for computing the OR, Jaccard and Forman curvatures for different real networks.
\label{table:RealWorldNetworks_running_time}}
\end{table}

\section{Conclusion}

We investigated a new network curvature metric inspired by the Jaccard coefficient, which we call the Jaccard curvature. We generalized the notion of Jaccard curvature and studied two Jaccard curvature metrics, JC and gJC. Theoretically, the gJC metric was shown to better approximate OR curvature for Erdos-Renyi graphs, compared to the JC metric. We conducted experiments with different classes of network models and real networks, and observed that gJC outperforms JC and Forman curvatures in approximating the OR curvature. Nonetheless, the JC curvature is easier to compute than gJC, and correlates moderately well with OR for positively curved or strongly clustered networks, suggesting that it could be used as a cheap proxy to the OR curvature for such special scenarios. The Forman curvature while being the cheapest to compute, shows weak correlation with OR curvature for many real networks. 

\bibliographystyle{IEEEtran}
\bibliography{ms}

\appendix
\externaldocument{ms.tex}
\externaldocument{curvature_setting.tex}
\externaldocument{theoretical_results.tex}

\section{Proof of Theorem \ref{thm:MainTheorem}}
\label{sec:Lemma_proofs}

\subsection{Preliminaries}

Since $\eqref{eq:mJCA_finalExpression}$ consists of ratios of rvs, we find it helpful to state a result which ensures mean convergence of ratio of rvs to the ratio of their means.

\subsubsection{A probabilistic result on ratios of rvs}
First we state a basic result which will help us investigate the asymptotic behavior of the gJC curvature.
\begin{lemma}
Consider the following sequences of rvs $\{N_n, n=1,2,\ldots \}$ and $\{D_n, n=1,2,\ldots \}$. 
If $\bE{N_n} \xrightarrow{n} c_1$ and $\bE{D_n} \xrightarrow{n} c_2$, then $\left| \bE{\frac{N_n}{D_n}} - \frac{c_1}{c_2} \right| \xrightarrow{n} 0$ as $n \to \infty$, provided $Var(D_n) \xrightarrow{n} 0$ as $n \to \infty$ and there exists finite constant $c$ such that $\left| \frac{N_n}{D_n} \right|  < c$ almost surely.
\label{lemma:ProbabilisticResult}
\end{lemma}
\myproof
For $n=1,2,\ldots$, 
\begin{align}
\bE{ \frac{N_n}{D_n} } = \bE{ \frac{N_n}{D_n} \1{  |D_n -c_2| < \epsilon} } + \bE{ \frac{N_n}{D_n} \1{  |D_n -c_2| > \epsilon} }.
\label{eq:LemmaER_expansion}
\end{align}
The latter term in \eqref{eq:LemmaER_expansion} yields
\begin{align}
\bE{ \frac{N_n}{D_n} \1{  |D_n -c_2| < \epsilon}^c }
&\leq c \bP{ |D _n - c_2|  >  \epsilon } 
\nonumber \\
&\leq c \frac{Var(D_n)}{\epsilon ^2}
\xrightarrow{n} 0
\label{eq:LemmaER_term2}
\end{align}
by using Chebyshev's inequality~\cite{grimmett2001probability}.
Using \eqref{eq:LemmaER_term2} in \eqref{eq:LemmaER_expansion}, we obtain
\begin{align}
\left| \bE{\frac{N_n}{D_n}} - \frac{c_1}{c_2} \right| \to 0
\end{align}
as $n \to \infty$ and $\epsilon \to 0$.
\myendpf

\subsubsection{Mean asymptotics on the sets of Separate nodes}

We aim to find the asymptotic behavior of $gJC(i,j)$ for an edge $(i,j)$.

Next, we state results on the asymptotics of the sets $S_{n,i} ^{(1)}$, $S_{n,i} ^{(2)}$ and $S_{n,i} ^{(3)}$ for $i=1,2$.

\begin{lemma}
For edge $(i,j) \in E_n$, we have 
\begin{align}
\bE{ S_{n,\ell} ^{(1)} } &\sim n p_n (1-p_n), \mbox{ when } \ p_n \to p \nonumber \\
 &\sim n p_n, \mbox{ when } \ n p_n ^2 \to \infty \nonumber \\
&\sim o(np_n), \mbox{ when } \ n p_n ^2 \to 0 \label{eq:S_1_ScalingResult}
\end{align}
for $\ell = i,j$, which also implies
\begin{align}
\bE{\left( S_{n,i} ^{(2)} + S_{n,j} ^{(2)} \right) + 
\left( S_{n,i} ^{(3)} + S_{n,j} ^{(3)} \right) } &\sim o(n p_n) , \ \mbox{ when } n p_n ^2 \to \infty \mbox{ or }  p_n \to p
\nonumber \\
& \sim 2 n p_n, \ \mbox{ when } n p_n ^2 \to 0.
\label{eq:S_2,S_3_ZeroScaling}
\end{align}
\label{lemma:S1-Result}
\end{lemma}
\myproof
Observe that,
\begin{align}
S_{n,1} ^{(1)} &= \{ k \in \cN _{n,1}(1,2) \ | \ \min _{\ell \in \cN _{n,2}(1,2)} d(k,\ell) = 1 \}
\nonumber \\
& = \{ k \in V_n \ | \ k \sim 1, k \nsim 2 \,\mbox{ and } \exists \ell \in \cN _{n,2}(1,2) \ : \ k \sim \ell \}
\nonumber \\
&= \sum _{k \in V_n} \1{k \sim 1, k \nsim 2} \left( 1 - 
\prod _{\ell \in V_n \setminus k} \left( \1{ \ell \sim 2 , \ell \nsim k } + \1{\ell \nsim 2} \right) \right).
\label{eq:S_1^1_expressions}
\end{align}

By using the iid property of edges in an Erdos Renyi graph, we obtain from \eqref{eq:S_1^1_expressions},
\begin{align}
\bE{S_{n,1} ^{(1)}} &= n p_n (1-p_n) \left( 1 - \left( p_n(1-p_n) + (1-p_n) \right) ^{n-1} \right)
\nonumber \\
&= np_n(1-p_n) \left( 1 - \left( 1 - p_n ^2 \right) ^{n-1} \right).
\label{eq:ES_1^1_expression}
\end{align}
Note that as $n \to \infty$ and $p_n \to 0$, $\bE{S_{n,1} ^{(1)}} \sim n p_n \left( 1 - e ^{ - n p_n ^2 } \right) $ and if $p_n \to p$, $\bE{S_{n,1} ^{(1)}} \sim n p (1-p)$. 
Therefore, first part of the lemma, \eqref{eq:S_1_ScalingResult} follows.
From \eqref{eq:SeparateVertices_asymp}, we note that $\bE{ S_n(i,j) } \sim 2 n p_n$. However, 
when $n p_n ^2 \to \infty$, $\bE{ S_{n,1} ^{(1)} + S_{n,2} ^{(1)} } \sim 2 n p_n$, implying that 
all separate vertices are actually in $\cS_{n,1} ^{(1)} \cup \cS_{n,2} ^{(1)}$. Thus the second part of the lemma follows as well.
\myendpf

Next, we study the properties of $S^{(2)}$.

\begin{lemma}
For edge $(i,j) \in E$, we have
\begin{align}
\bE{S_{n,\ell} ^{(2)}} &\sim n p_n, \ n ^2 p_n ^3 \to \infty, n p_n ^2 \to 0 
\nonumber \\
& \sim o(n p_n), \ n ^2 p_n ^3 \to 0
\label{eq:S2_scalingResult}
\end{align}
for $\ell =(i,j)$, which also implies
\begin{align}
\bE{ \left( S_{n,i} ^{(3)} + S_{n,j} ^{(3)} \right)  } &\sim o (n p_n), \ n ^2 p_n ^3 \to \infty
\nonumber \\
&\sim 2 n p_n, \ n ^2 p_n ^3 \to 0, n p_n \to \infty.
\label{eq:S3_ZeroScalingResult}
\end{align}
\label{lemma:S2_S3_Result}
\end{lemma}
\myproof
Observe that
\begin{align}
S_1 ^{(2)} &= 
\{ k \in \cN _{n,1}(1,2) \ | \ \min _{\ell \in \cN _{n,2}(1,2)} d(k,\ell) = 2 \}
\nonumber \\
&=
\{ k \in \cN_{n,1}(1,2) \setminus (\cC \cup \cS_{n,1}^{(1)}) \ | \ \exists \ell,m \in V_n
 \mbox{ s.t. } \ell \sim 2, \ell \nsim 1, \ell \sim m, m \sim k \}.
 \label{eq:S2_expression1}
\end{align}
From \eqref{eq:S_2,S_3_ZeroScaling}, we note that we only need to consider the scaling where 
$np_n ^2 \to 0$. Therefore, we can drop the requirement of $k \notin S_{n,i} ^{(1)}$. 
Continuing from \eqref{eq:S2_expression1}, 
\begin{align}
S_{n,1} ^{(2)} &= \sum _{k \in V_n} \1{k \sim 1, k \nsim 2} 
\nonumber \\
& \times
\left( 
1 - \prod _{\ell \in V_n \setminus \{1,2,k\}} \prod _{m \in V_n \setminus \{ 1,2,k,\ell \}}
\left(  \1{ \ell \sim 2, k \sim m, m \nsim \ell } + \1{\ell \sim 2, k \nsim m }
+ \1{\ell \nsim 2} \right) \right).
\nonumber
\end{align}  
Using the independence of the involved rvs $\{k \sim 1 \}$,$\{k \nsim 2\}$, $\{\ell \sim 2 \}$, 
$\{k \sim m \}$ and $\{m \nsim \ell \}$, we obtain
\begin{align}
\bE{S_{n,1} ^{(2)} }  &= n p_n (1-p_n) \left( 1 - \left( p_n ^2(1-p_n) +p_n(1-p_n) +(1-p_n) 
\right) ^{(n-3)(n-4)}
\right)
\nonumber \\
&= n p_n (1-p_n) \left( 1 - p_n ^3 \right) ^{ (n-3)(n-4) }.
\label{eq:S2_expression2}
\end{align}
Note that as $n \to \infty$, $\bE{S_{n,1} ^{(2)}} \sim n p_n e ^{- n^2 p_n ^3}$. Therefore,
first part of the lemma follows.
This implies that in the range $n ^2 p_n ^3 \to \infty, n p_n ^2 \to 0$, all separate vertices are actually in $\cS_{n,1} ^{(2)} \cup \cS_{n,2} ^{(2)}$. For the range $n ^2 p_n ^3 \to 0, n p_n \to \infty$, since $\bE{S_n(1,2)} \sim 2 n p_n$, we must have all the separate nodes in $\cS_{n,1} ^{(3)} \cup \cS_{n,2} ^{(3)}$.
\myendpf

\subsubsection{Variance asymptotics}

In this section we analyze the variance asymptotics of the set of neighbor nodes.

\begin{lemma}
For edge $(i,j)$ in $E$, we have
\begin{equation}
Var(N_n(i,j)) \sim 2 n p_n.
\end{equation}
\label{lemma:Variance_Result}
\end{lemma}
\myproof
 Observe that, 
\begin{equation}
N_n(1,2) = n - \sum _{k \in V_n \setminus \{1,2\}} \1{k \nsim 1} \1{k \nsim 2}.
\end{equation}
Set $\chi _k = \1{k \nsim 1} \1{k \nsim 2}$. It follows that
\begin{align}
Var (N_n(1,2)) = \sum _{k \in V_n \setminus \{1,2\}} Var (\chi _k) + 
\sum _{i \neq j} Cov(\chi _i,  \chi _j).
\end{align}
By independence of links in an ER graph, $Cov(\chi _i,  \chi _j) = 0$, and 
\begin{align}
Var(\chi _k) &= \bE{\1{k \nsim 1} \1{k \nsim 2}} - 
\left( \bE{\1{k \nsim 1}} \bE{\1{k \nsim 2}} \right) ^2
\nonumber \\
&= (1-p_n) ^2 - (1-p_n) ^4
\nonumber \\
&= (1-p_n) ^2 (2 p _n - p_n ^2).
\end{align}
Therefore, 
\begin{equation}
Var(N_n(1,2)) \sim 2 n p_n. 
\end{equation}
\myendpf


\label{sec:VarianceAsymptotics}

\section{Complexity analysis for General Graph}
\label{sec:General_complexity}

We introduce some notation: Let $n$, $m$ and $\Delta$ denote the number of nodes, number of edges and the maximum degree in a graph respectively. Let $d_x$ denote the degree of node $x$, and $D_x$ the sum of the degrees of all the neighbors of $x$.

\begin{lemma}
Total cost of computing $S^{(0)}$ for all the edges of the graph is $2\sum_{v\in V}d_v^2$
\end{lemma}
\myproof
As in Lemma~\ref{lemma:S0}, merge-sort $N(i), N(j)$ will cost $d_i+d_j$, so the total cost will be  $\sum_{(i,j)\in E}(d_i+d_j)$. It is easy to see that each $d_i$ appears $d_i$ time in the summation, therefore the total cost is $2\sum_{v\in V} d_v^2$
\myendpf

\begin{lemma}
Total cost of computing $S^{(1)}$ for all the edges of the graph is $2\sum_{v\in V}d_vD_v$
\end{lemma}
\myproof
Following a process similar to that for the proof of Lemma~\ref{lemma:S1}, the cost of calculating $S^{(1)}(i,j)$ is 
\[\sum_{x\in N(i)} (d_x+d_j) +\sum_{y\in N(j)}(d_y+d_i)=\sum_{x\in N(i)}d_x +\sum_{y\in N(j)}d_y+ 2d_id_j=D_i+D_j+2d_id_j\]
The total cost is $\sum_{(i,j)\in E} (D_i+D_j+2d_id_j)$. Notice that 
\[\sum_{(i,j)\in E} (D_i+D_j)=\sum_{v\in V} d_vD_V\]
and 
\[\sum_{(i,j)\in E} 2d_id_j=\sum_{i\in V} \sum_{j\in N(i)} d_id_j=\sum_{i\in V} d_i(\sum_{j\in N(i)} d_j)=\sum_{i\in V} d_iD_i\]

Therefore the total cost is $2\sum_{v\in V} d_vD_v$
\myendpf

\begin{lemma}
Total cost of computing $S^{(2)}$ and $S^{(3)}$  for all the edges of the graph is $\sum_{v\in V} (d_v+1)D_v$

\end{lemma}

\myproof
The proof is similar to that for Lemma~\ref{lemma:S2} The preprocessing of BFS costs $\sum_{v\in V} D_v$. The cost for determining each $S^{(2)}(i,j)$ or $S^{(3)}(i,j)$ is
\[\sum_{x\in N(i)} d_j +  \sum_{y\in N(j)}d_i=2d_id_j\]

So the total cost is $\sum_{(i,j)\in E} 2d_id_j=\sum_{v\in V}d_vD_V$, overall cost is $\sum_{v\in V} (d_v+1)D_v$
\myendpf

Using the lemmas stated above we obtain the following result.
\begin{theorem}
The complexity of computing JC for a general graph is $O(2\sum_{v\in V}d_v^2)$, for 
gJC is $O(\sum_{v\in V} (d_v+1)D_v)$, and for OR is $O(\sum_{(i,j)\in E}(d_i+d_j+d_i*d_j)(d_i+d_j)\log^2 (d_i+d_j))$.
\end{theorem}








\end{document}